\documentclass[10pt,tightenlines,eqsecnum,floats,aps,nofootinbib,prd]{revtex4-2}
\usepackage{amsmath,amssymb,amsfonts,amsthm,amscd}
\usepackage{graphicx}
\usepackage{amsmath}
\usepackage{amssymb}
\usepackage{graphicx}
\usepackage{cancel}
\usepackage{slashed}
\usepackage{mathtools}
\usepackage{tcolorbox}
\usepackage{tikz-cd}
\usepackage{tikz}
\usepackage{mathrsfs}
\usepackage{relsize}
\usepackage{natbib}
\usepackage[margin=1in]{geometry}
\usepackage{hyperref}
\usepackage[normalem]{ulem}

\usetikzlibrary{shapes.geometric, arrows}
\tikzstyle{startstop} = [rectangle, rounded corners, minimum width=3cm, minimum height=1cm,text centered, draw=black, fill=red!30]
\tikzstyle{arrow} = [thick,->,>=stealth]
\begin{document}
\title{Dynamical Similarity in Higher - Order Classical Symplectic Systems}
\author{Callum Bell}
\email{c.bell8@lancaster.ac.uk}
\affiliation{Department of Physics, Lancaster University, Lancaster UK}

\author{David Sloan}
\email{d.sloan@lancaster.ac.uk}
\affiliation{Department of Physics, Lancaster University, Lancaster UK}

\begin{abstract}
    Many theories of physical interest, which admit a Hamiltonian description, exhibit symmetries under a particular class of non - strictly canonical transformation, known as dynamical similarities. The presence of such symmetries allows a reduction process to be carried out, eliminating a single degree of freedom from the system, which we associate with an overall scale. This process of `contact reduction' leads to theories of a frictional nature, in which the physically - observable quantities form an autonomous subsystem, that evolves in a predictable manner. We demonstrate that this procedure has a natural generalisation to theories of higher order; detailed examples are provided, and physical implications discussed.
\end{abstract}
\maketitle

\section{Introduction}\label{Sec:Intro}

In general, it is customary to adopt the viewpoint that isolated systems should be afforded a privileged status within the framework used to describe physical law \cite{cuffaro2021open}. We confer particular importance to such systems, deeming them more fundamental than their open counterparts. Indeed, when describing systems exhibiting dissipative effects, we choose to subsume this within our closed - system approach, treating open systems as part of a larger entity, comprising the original system coupled to some `environment'. Matter, energy, and information may be freely exchanged between the two subsystems, but the full `system plus environment' complex is still considered to be closed.\\

It was argued by Leibniz that the universe as a whole is the only truly closed system \cite{leibniz2000leibniz}; consequently, if we dogmatically insist that closed systems are fundamental, we must logically conclude that the universe is the only fundamental entity that exists \cite{schaffer2013jonathan}. When studying symmetry - reduced solutions of the Einstein field equations, particularly within the context of Friedmann - Lemaître - Robertson - Walker (FLRW) cosmologies, we do indeed treat the universe as a closed system; however, as highlighted in \cite{smeenk2013philosophy}, this is little more than an idealisation used to reduce calculational complexity. A particularly noteworthy example of this, is the use of the scale factor, which, despite its pivotal role in the geometrical description offered by General Relativity, is generally taken to carry no physical meaning \cite{sloan2021new}. Moreover, it is possible to excise completely this unphysical degree of freedom from our cosmological ontology, leaving a purely relational description.\\

It has been proposed that, when constructing a theory, we should first find a framework which amply admits the phenomena we wish to describe, before imposing successively more restrictive conditions, so as to reduce the amount of superfluous structure present \cite{ismael2021symmetry}. In the current context, we posit that the notion of superfluous structure may be cast as a precise set of dynamical criteria, captured by the so - called Principle of Essential and Sufficient Autonomy (PESA) \cite{gryb2021scale}. We seek a minimal ontology that closes the dynamical algebra of observables, and any structure which is not contained within this ontology is deemed surplus; since the only degrees of freedom accessible via empirical methods are dimensionless ratios, we are pushed towards a relational description of nature.\\

This is quite a radical step, to claim that we should seek a description of reality based entirely on relational terms, and we ought to provide some additional motivation. As argued by Poincar\'e, a hypothetical scenario, in which all distances within our universe were suddenly increased to twice their original values, would be entirely indistinguishable from the initial configuration. This follows from the fact that our ability to deduce any such changes of scale relies upon a set of measuring instruments, which have undergone an identical doubling in size \cite{poincare2003science}.\\

Detractors of this viewpoint, such as Galileo, and later Delboeuf \cite{galilei1914dialogues,delboeuf1894dimensions}, reason that this argument is flawed, since fundamental parameters, such as Newton's gravitational constant, may be used to infer that a change of scale has occurred, through the fact that biological systems, for example, would be unable to support their enlarged skeletal structures. However, we argue that the fundamental constants of nature are necessarily deduced through experiment and observation: for the case of Newton's gravitational constant, for example, this required the use of the torsion balance, pioneered by Cavendish \cite{cavendish1798xxi}. Consequently, it is entirely logical to consider that the artifice producing this hypothetical change of scale should also act on the parameters of our theory, thereby restoring our ability to provide a wholly relational description of nature. This issue is further explored in section (\ref{Sec:Couplings}).\\

Despite the fact that the Lagrangian and Hamiltonian formulations of classical mechanics have been studied extensively throughout history \cite{lachieze2014historical,brizard2014introduction,mohallem2024lagrangian,zain2019techniques}, it was not until relatively recently, with the work of mathematicians such as Souriau and Tulczyjew \cite{souriau1970structure,souriau1986structure,tulczyjew1976sous,wlodzimierz1976sous}, that these theories were placed on a geometrical footing, allowing the powerful tools of differential geometry to be applied to the study of mechanical systems \cite{guggenheimer2012differential}. In seeking a relational description of the laws governing our universe, we shall find that we are forced to work with manifolds on which phase space volumes are not conserved, and mechanical energy lost. In this way, by arguing that a minimalist approach should be taken, when attempting to formulate physical law, we are (indirectly) advocating that open systems should in fact be granted privileged status as more fundamental.\\

In this article, we commence by presenting an introduction to higher - order geometrical mechanics, in both the Lagrangian and Hamiltonian settings. We favour the geometrical approach over the more familiar variational formulation, as our ultimate goal will be to extend our formalism to those theories whose dynamical objects are fields. This, while possible, is more difficult to do using traditional variational methods. For readers unfamiliar with the geometrical methods described, we shall present a running example, in which the underlying concepts may be understood, in a more mathematically - amenable setting. We also include an appendix, in which the variational approach is shown to produce equivalent results.\\

In section (\ref{Sec:ScalingSym}), we illustrate the mechanism by which the changes in scale discussed above may be implemented, before exploring how symmetry under these transformations provides a natural way in which to eliminate overall scale from our ontology. Original results, on the generalisation of this reduction process to higher - order theories are then presented in section (\ref{Sec:HigherOrderContRed}). Finally, we conclude with a number of examples, discussing both the utility and physical significance of our findings.

\section{Higher - Order Symplectic Mechanics}\label{Sec:HigherOrderSymplecticMech}

\subsection{Lagrangian Formalism}\label{Subsec:LagrangeSymp}
A geometric formulation of theories of higher order requires the introduction of a class of manifold known as \textit{higher - order tangent bundles} \cite{de1989connections}. In general, if $Q$ is an $n$ - dimensional smooth manifold, with bundle structure $\pi: TQ\rightarrow Q$, we define the $n(k+1)$ - dimensional space $T^kQ$ as the collection of all $k$ - jets from $0\in \mathbb{R}$ to $Q$ \cite{saunders1989geometry,kupershmidt2006geometry}, which we express as
\begin{equation*}
    T^kQ\equiv J_0(\mathbb{R},Q) = \bigcup\limits_{q\,\in\,Q} \{j^k_0\phi\;|\; \phi:I\subset\mathbb{R}\rightarrow Q\}
\end{equation*}
in which $j^k_0\phi$ denotes the $k$ - jet of the curve $\phi: I\subset \mathbb{R}\rightarrow Q$. \\

The manifold $T^kQ$ admits the following natural projections
\begin{equation*}
    \begin{split}
        \rho^k_r:T^kQ&\rightarrow T^rQ \quad\quad\quad \rho^k_0 \equiv\beta^k: T^kQ \rightarrow Q\\
        j^k_0\phi &\mapsto j^r_0\phi \quad\quad\quad\quad\quad\quad\quad\quad \,j^k_0\phi \mapsto \phi(0)
    \end{split}
\end{equation*}
Local coordinates on $T^kQ$ are constructed using a coordinate chart $(U,\varphi)$ of $Q$, and a curve $\gamma: I\subset\mathbb{R}\rightarrow Q$ with $\gamma(0)\in U$. We then define $\gamma^i = \varphi^i\circ \gamma$, and local coordinates on $(\beta^k)^{-1}(U)\equiv T^kU$ are $(q^i,q^i_1,\,\cdots\,,q^i_k)$, where
\begin{equation*}
    q^i = \gamma^i(0) \quad\quad\textrm{and}\quad\quad q_{\alpha}^i=\frac{d^{\,\alpha}\gamma^i}{dt^{\alpha}}\biggr|_{t=0}
\end{equation*}
Writing $q^i\equiv q_0^i$, we collectively denote local coordinates on $T^kU$ by $(q^i_{\alpha})$. Physically, we identify $q^i$ as a generalised coordinate, with $q^i_{\alpha}$ its time derivative of order $\alpha$.\\

In contrast to the case of first - order mechanics, the Lagrangian is now \textit{not} defined on the same space as the remaining dynamical objects \cite{koszul2019introduction,libermann2012symplectic,da2001lectures}. In particular, the Lagrangian is a function $\mathcal{L}\in C^{\infty}(T^kQ)$, whereas the $k^{\textrm{th}}$ order Lagrangian 1 and 2 - forms, together with the energy function, are defined over the space $T^{2k-1}Q$. We shall see throughout that this seemingly trivial feature has extremely important implications for the contact reduction process. In light of this, we are motivated to introduce the \textit{Tulczyjew total derivative} \cite{prieto2014geometrical, de2021higher}, which is a map $d_T:C^{\infty}(T^rQ)\rightarrow C^{\infty}(T^{r+1}Q)$, whose action on a function may be expressed locally as
\begin{equation}\label{Eq:Tul}
    d_Tf = \sum_{\alpha=0}^rq^i_{\alpha+1}\frac{\partial f}{\partial q^i_{\alpha}}
\end{equation}
In more physical terms, we should consider the derivative $d_T$ to be the usual total time derivative, adapted to the underlying bundle structure.\\

On the manifold $T^{2k-1}Q$, we have coordinates $q^i_{\alpha}$, with $0\leqslant \alpha \leqslant 2k-1$; however, we may also consider the set $(q^i_{\alpha},\widehat{p}^{\;\alpha}_i)$, with $0\leqslant \alpha\leqslant k-1$, where
\begin{equation}\label{Eq:jom}
    \widehat{p}^{\;r-1}_i = \sum_{\alpha=0}^{k-r} (-1)^{\alpha} d^{\,\alpha}_T\left(\frac{\partial \mathcal{L}}{\partial q^i_{r+\alpha}}\right)
\end{equation}
(for $1\leqslant i\leqslant n$ and $1\leqslant r\leqslant k$). The $\widehat{p}^{\;\alpha}_i$ are known as \textit{generalised Jacobi - Ostrogradsky momenta}, and satisfy the recursion relation
\begin{equation}\label{Eq:rec}
    \widehat{p}^{\;r-1}_i = \frac{\partial \mathcal{L}}{\partial q^i_r} - d_T(\widehat{p}^{\;r}_i)
\end{equation}
In these coordinates, the Lagrangian 1 and 2 - forms $\theta_{\mathcal{L}}$ and $\omega_{\mathcal{L}} = -\,d\theta_{\mathcal{L}}$ are given by\footnote{Note that we have suppressed sums over both indices $1\leqslant i\leqslant n$ and $0\leqslant\alpha\leqslant k-1$. We shall continue to do so, except for when explicit summation provides additional clarity.}
\begin{equation}\label{Eq:theta}
    \theta_{\mathcal{L}} = \widehat{p}^{\;\alpha}_i\,dq^i_{\alpha}\quad\quad\quad\quad \omega_{\mathcal{L}} = dq^i_{\alpha}\wedge d\widehat{p}^{\;\alpha}_i
\end{equation}
Additionally, the energy function $E_{\mathcal{L}}$ may be expressed as 
\begin{equation}\label{Eq:energy}
    E_{\mathcal{L}} = \sum_{\alpha=0}^{k-1} \widehat{p}^{\;\alpha}_i q^i_{\alpha+1} - \mathcal{L}
\end{equation}
We refer to the pair $(T^{2k-1}Q,\mathcal{L})$ as a \textit{$k^{\textrm{th}}$ order Lagrangian system}, and further specify this system to be \textit{regular} \cite{de1992inverse} if
\begin{equation*}
    \textrm{det}\left(\frac{\partial^2 \mathcal{L}}{\partial q^i_{k}\partial q^j_k}\right) (j^{2k-1}_0\phi)\neq 0 \quad\textrm{for all points}\; j^{2k-1}_0\phi \in T^{2k-1}Q
\end{equation*}
As in the first - order case, the equations of motion are found introducing a vector field $X_{\mathcal{L}}\in \mathfrak{X}^{\infty}(T^{2k-1}Q)$, such that
\begin{eqnarray}\label{Eq:geomEL}
    i_{\scriptscriptstyle X_{\mathcal{L}}}\omega_{\mathcal{L}} = dE_{\mathcal{L}}
\end{eqnarray}
For systems of physical interest, integral curves of $X_{\mathcal{L}}$ may always be written as the canonical lifting (or prolongation) of curves $\phi_{\mathcal{L}}:\mathbb{R}\rightarrow Q$ to $T^{2k-1}Q$, and it is these curves $\phi_{\mathcal{L}}$ which correspond to physical trajectories of the system \cite{de2021higher}. If we express $\phi_{\mathcal{L}}$ in local coordinates as $\phi_{\mathcal{L}}(t) = (q^i(t))$, the prolongation $j_0^{2k-1}\phi_{\mathcal{L}}$ is given by
\begin{equation*}
    j_0^{2k-1}\phi_{\mathcal{L}}(t) = \left(q^i(t),\frac{dq^i}{dt}\biggr|_t ,\,\,\cdots\,\,\frac{d^{2k-1}q^i}{dt^{2k-1}}\biggr|_t\right)
\end{equation*}
The curve $\phi_{\mathcal{L}}$ then satisfies the $k^{th}$ \textit{order Euler - Lagrange equations}, which are given by
\begin{equation}\label{Eq:EL}
    \sum_{\alpha=0}^k (-1)^{\alpha} \frac{d^{\,\alpha}}{dt^{\alpha}}\left(\frac{\partial \mathcal{L}}{\partial q^i_{\alpha}}\right)\biggr|_{j^{2k-1}_o\phi_{\mathcal{L}}} = \,0
\end{equation}
When considering theories containing higher - order derivatives, it is important to note that boundary conditions must be treated with significantly more care. Typically, for first - order theories, we carry out a variational calculation, subject to the requirement that each of the variations $\delta q^i$ vanish on the boundary; in this way, any total derivative added to the Lagrangian is inconsequential from the perspective of the physical dynamics. This fact leaves us free to `discard' boundary terms in a somewhat careless manner.\\

By contrast, for a theory containing derivatives of up to order $k$, there are, in addition to each of the $\delta q^i$, the variations $\delta q^i_{\alpha}$ of the coordinate derivatives. In this case, we should fix each of these variations on the boundary for $\alpha=0,\,\cdots\,,k-1$; indeed, failure to do so alters the physical solution space of the resulting theory \cite{fatibene2011boundarytermshigherorder}. In general, boundary conditions should be chosen to faithfully reflect the physical phenomena to be described by the model of interest \cite{yao2020generalized}; such a choice often requires the introduction of additional surface terms, in order to impose these conditions, and simultaneously ensure the well - posedness of the variational principle. A particularly noteworthy example of this arises in the study of higher - order theories of gravity \cite{Guarnizo_2010,barth1985fourth,muller1990gravity}, in which surface terms are added, similar to those of Gibbons - Hawking - York.\\

In the current context, we are interested in the study of higher - order mechanical systems which exhibit dynamical similarities; such systems are, of course, a subset of the space of higher - derivative theories, and so are naturally subject to the same class of considerations regarding boundary conditions as outlined above. In the interest of minimising additional complications, which direct attention away from our main focus - the presentation of a symmetry reduction framework for higher - order theories - we shall restrict ourselves to Dirichlet conditions, such that, on the boundary, we impose $\delta q^i_{\alpha}=0$ for $\alpha=0,\cdots,k-1$. While this may seem somewhat unphysical, we emphasise that it is merely for the purpose of simplicity, and that no unexpected subtleties arise within the procedure outlined in section (\ref{Sec:HigherOrderContRed}), if one considers alternative boundary conditions.
\subsection{The Legendre - Ostrogradsky Map}\label{Subsec:LegSymp}
Higher - order Hamiltonian mechanics takes place on the cotangent bundle $T^*(T^{k-1}Q)$, which is endowed with the canonical 1 and 2 - forms $\theta$ and $\omega$. The Legendre - Ostrogradsky map is the generalisation of the Legendre transform from $TQ$ to $T^*Q$ \cite{libermann2012symplectic,colombo2016regularity,arnolʹd1990symplectic}. In particular, it is a map $\mathcal{FL}:T^{2k-1}Q\rightarrow T^*(T^{k-1}Q)$ which satisfies
\begin{equation*}
    \mathcal{FL}^*\,\theta = \theta_{\mathcal{L}} \quad\quad\quad\quad \mathcal{FL}^*\,\omega = \omega_{\mathcal{L}}
\end{equation*}
We take coordinates on $T^*(T^{k-1}Q)$ to be $(q^i_{\alpha},p^{\alpha}_i)$, for $0\leqslant \alpha\leqslant k-1$, where $\mathcal{FL}^*p^{\alpha}_i = \widehat{p}^{\;\alpha}_i$. As such, it follows that
\begin{equation}\label{Eq: theta2}
    \theta = p^{\alpha}_i\,dq^i_{\alpha}\quad\quad\quad\quad \omega= dq^i_{\alpha}\wedge dp^{\alpha}_i    
\end{equation}
provide local coordinate expressions for the canonical 1 and 2 - forms on $T^*(T^{k-1}Q)$.
\subsection{Hamiltonian Formalism}\label{Subsec:HamiltonianSymp}
If the Legendre - Ostrogradsky map is a global diffeomorphism, there exists a unique and globally - defined Hamiltonian function $\mathcal{H}\in C^{\infty}(T^*(T^{k-1}Q))$ such that $\mathcal{FL}^*\mathcal{H}=E_{\mathcal{L}}$ \cite{colombo2016regularity}. Additionally, we have a vector field $X_{\mathcal{H}}\in \mathfrak{X}^{\infty}(T^*(T^{k-1}Q))$ satisfying
\begin{equation}\label{Eq:ham}
    i_{\scriptscriptstyle X_{\mathcal{H}}} \omega= d\mathcal{H}
\end{equation}
In local coordinates $(q^i_{\alpha},p^{\alpha}_i)$ on $T^*(T^{k-1}Q)$, the Hamiltonian is given by
\begin{equation}\label{Eq:hamloc}
    \mathcal{H}= \sum_{\alpha=0}^{k-2} q^i_{\alpha+1}\,p^{\alpha}_i + (\mathcal{FL}^{-1})^*\,q^i_{k}\,p_i^{k-1} -(\mathcal{FL}^{-1})^*\,\mathcal{L}
\end{equation}
Integral curves of the Hamiltonian vector field $X_{\mathcal{H}}$ correspond to physical trajectories of the system, and so if we express such an integral curve in local coordinates as $\psi_{\scriptscriptstyle \mathcal{H}}(t) = (q^i_{\alpha}(t)\,,\,p^{\alpha}_i(t))$, we find that
\begin{equation}\label{Eq:HE}
    \dot{q}^i_{\alpha} = \frac{\partial \mathcal{H}}{\partial p^{\alpha}_i}\biggr|_{\psi_{\scriptscriptstyle\mathcal{H}}(t)} \quad\quad\quad \dot{p}_i^{\alpha} = -\, \frac{\partial \mathcal{H}}{\partial q_{\alpha}^i}\biggr|_{\psi_{\scriptscriptstyle\mathcal{H}}(t)} 
\end{equation}
which are the local form of the $k^{\textrm{th}}$ order Hamilton equations.
\subsection{Example: The Pais - Uhlenbeck Oscillator}\label{Subsec:PUExample}
As outlined in the introduction, we anticipate that many a reader with a strictly physics background may find the geometrical concepts we have presented to be highly unfamiliar. As such, in an attempt to facilitate understanding, we shall consider one of the simplest examples of a higher - order system: the Pais - Uhlenbeck oscillator. For details, comments, and a more in - depth analysis of this model, we refer to \cite{bolonek2005hamiltonian,smilga2009comments,andrzejewski2014hamiltonian}, for example.\\

Traditionally, the configuration space $Q$ of the Pais - Uhlenbeck oscillator is parameterised via a single generalised coordinate $q$, and so is of dimension one. However, in order to provide a satisfactory discussion of the symmetry reduction process, it will be advantageous to have a second generalised coordinate. In this way, we will be able to distinguish between scaling variables, and those parameters which remain unchanged. For this reason, we shall promote the constant frequency $\omega$ of the oscillator to a velocity $\dot{\theta}$, corresponding to the coordinate $\theta$. The Lagrangian of the system then contains up to second derivatives, and so is a function on $T^2Q$. Coordinates on this space are taken to be $(q_0,q_1,q_2,\theta_0,\theta_1)$, which we shall identify with the more familiar $(q,\dot{q},\ddot{q},\theta,\dot{\theta})$. As such, the Lagrangian for this oscillator is
\begin{equation}\label{Eq:PUL}
    \mathcal{L} = \frac{1}{2}\left(\dot{q}^2 - q^2\dot{\theta}^2 - \lambda\ddot{q}^2\right)
\end{equation}
in which $\lambda$ is real constant, characteristic of the oscillator. From (\ref{Eq:EL}), we see that the equations of motion for the variables $q$ and $\theta$ are simply
\begin{equation*}
    \lambda \,q^{(4)} + \ddot{q} + q\dot{\theta}^2 = 0 \quad\quad\quad\textrm{and}\quad\quad\quad q^2 \ddot{\theta} + 2 q \dot{q} \dot{\theta} = 0
\end{equation*}
On the space $T^{2k-1}Q = T^3Q$, we introduce the Jacobi - Ostrogradsky momenta $\widehat{p}_q^{\;0}$, $\widehat{p}_q^{\;1}$, and $\widehat{p}^{\;0}_{\theta}$, where
\begin{equation*}
    \widehat{p}_q^{\;1} = \frac{\partial \mathcal{L}}{\partial \ddot{q}} = -\,\lambda\ddot{q} \quad\quad\quad \widehat{p}_q^{\;0} = \frac{\partial \mathcal{L}}{\partial \dot{q}} - d_T(\widehat{p}_q^{\;1}) = \dot{q}+\lambda\dddot{q} \quad\quad\quad \widehat{p}^{\;0}_{\theta} = \frac{\partial\mathcal{L}}{\partial \dot{\theta}} = -\, q^2\dot{\theta}
\end{equation*}
The energy function is calculated according to (\ref{Eq:energy}); however, recall that we take coordinates on $T^3Q$ to be $(q,\dot{q},\theta,\widehat{p}_q^{\;0},\widehat{p}_q^{\;1},\widehat{p}^{\;0}_{\theta})$. Thus, eliminating $\dot{\theta}$, $\ddot{q}$, and $\dddot{q}$ using the momenta calculated above, we find that
\begin{equation}\label{Eq:PUenergy}
    E_{\mathcal{L}} = \dot{q}\,\widehat{p}_q^{\;0} - \frac{1}{2}\left(\dot{q}^2 + \frac{1}{q^2}(\widehat{p}^{\;0}_{\theta})^2 + \frac{1}{\lambda}(\widehat{p}_q^{\;1})^2\right)
\end{equation}
For pedagogical reasons, let us instead express this energy function in terms of $q$, $\theta$, and their time derivatives:
\begin{equation}
    \label{Eq:PUELinq}
    E_\mathcal{L} = \frac{\dot{q}^2}{2} + \lambda \dot{q} \dddot{q} - \frac{q^2 \dot{\theta}^2}{2}-\frac{\lambda \ddot{q}^2}{2}
\end{equation}
In these same coordinates, the Lagrangian 2 - form $\omega_{\mathcal{L}}$ is given by
\begin{equation}
    \label{Eq:PUOmegaLinq}
    \omega_\mathcal{L} = dq \wedge d \dot{q} + \lambda\, dq \wedge d \dddot{q} \, - \lambda \, dq \wedge d\ddot{q} +2 q \dot{\theta}\, dq \wedge d \theta - q^2 d\theta\wedge d\dot{\theta}
\end{equation}
Although the use of the $\widehat{p}$ variables introduces an algebraic simplification, the form of $E_{\mathcal{L}}$ and $\omega_{\mathcal{L}}$ given above offers a more direct way to see how these object behave under rescalings of $q$, which leave $\theta$ invariant.\\

The Lagrange vector field introduced in (\ref{Eq:geomEL}) is given by
\begin{equation}
    \label{Eq:PUXLinq}
    X_\mathcal{L} = \dot{q} \frac{\partial}{\partial \dot{q}} + \ddot{q}  \frac{\partial}{\partial \dot{q}} + \dddot{q}  \frac{\partial}{\partial \ddot{q}} - \frac{\ddot{q}+q \dot{\theta}^2}{\lambda}  \frac{\partial}{\partial \dddot{q} } + \dot{\theta}  \frac{\partial}{\partial \theta } - \frac{2 \dot{q} \dot{\theta}}{q}  \frac{\partial}{\partial \dot{\theta}}
\end{equation}
and it is straightforward to verify that $i_{\scriptscriptstyle X_{\mathcal{L}}}\omega_{\mathcal{L}} = dE_{\mathcal{L}}$, as expected.\\

The Hamiltonian is easily found, replacing each $\widehat{p}^{\,\alpha}$ with its counterpart $p^{\alpha}$
\begin{equation}
    \mathcal{H} = \dot{q}\,p_q^0 - \frac{1}{2}\left(\dot{q}^2 + \frac{1}{q^2}(p^0_{\theta})^2 + \frac{1}{\lambda}(p_q^1)^2\right)
\end{equation}
from which the equations of motion
\begin{align*}
    \ddot{q} &= -\,\frac{1}{\lambda}p^1_q & \dot{p}^0_q &= -\,\frac{1}{q^3}(p_{\theta}^0)^2 & \dot{p}^0_{\theta} &= 0\\
    \dot{\theta} &= -\,\frac{1}{q^2}p^0_{\theta} & \dot{p}^1_q &= -p^0_q +\dot{q}
\end{align*}
immediately follow.

\section{Higher - Order Contact Mechanics}\label{Sec:HigherOrderContactMech}

\subsection{Lagrangian Formalism}\label{Subsec:LagrangeContact}
The higher - order analogue of the extended tangent bundle $TQ\times\mathbb{R}$, upon which we first study contact mechanics, is the space $T^kQ\times\mathbb{R}$, with local coordinates $(q^i_0,\,\cdots\,,q^i_k,z)$. Here, the $q^i_{\alpha}$ are simply the coordinates of $T^kQ$, as defined above, and $z$ is to be considered a parameter spanning $\mathbb{R}$.\\

Many of the results of the preceding sections may be carried over to the space $T^kQ\times\mathbb{R}$, with suitable modification. For example, the Tulczyjew derivative is replaced by the map $D_{\mathcal{L}}: C^{\infty}(T^rQ\times\mathbb{R})\rightarrow C^{\infty}(T^{r+1}Q\times \mathbb{R})$, which depends explicitly upon the choice of Lagrangian function $\mathcal{L}\in C^{\infty}(T^kQ\times\mathbb{R})$. We refer to $D_{\mathcal{L}}$ as the \textit{Lagrangian total derivative}; however, strictly speaking, $D_{\mathcal{L}}$ is not a derivative, since it does not satisfy the Leibniz rule. Its action on a function is given locally by
\begin{equation}\label{Eq:Lagtot}
    D_{\mathcal{L}}f = \sum_{\alpha=0}^rq^i_{\alpha+1}\,\frac{\partial f}{\partial q^i_{\alpha}} + \mathcal{L}\,\frac{\partial f}{\partial z} - f\,\frac{\partial \mathcal{L}}{\partial z}
\end{equation}
The generalised Jacobi - Ostrogradsky momenta $\widehat{p}^{\;r-1}_i$ on $T^{2k-1}Q\times\mathbb{R}$ are then calculated in an identical manner to their symplectic counterparts, exchanging $d_T$ for $D_{\mathcal{L}}$ :
\begin{equation}\label{Eq:contactmomenta}
    \widehat{p}^{\;r-1}_i = \sum_{\alpha=0}^{k-r} (-1)^{\alpha} D_{\mathcal{L}}^{\alpha}\left(\frac{\partial \mathcal{L}}{\partial q^i_{r+\alpha}}\right)
\end{equation}
It is then evident that the contact analogue of the recursion relation (\ref{Eq:rec}) is simply
\begin{equation}\label{Eq:rec2}
    \widehat{p}^{\;r-1}_i = \frac{\partial \mathcal{L}}{\partial q^i_r}-D_{\mathcal{L}}(\widehat{p}^{\;r}_i)
\end{equation}
On the space $T^{2k-1}Q\times \mathbb{R}$, in addition to the Lagrangian 1 and 2 - forms, we also require a contact form $\eta_{\mathcal{L}}\in \Omega^1(T^{2k-1}Q\times\mathbb{R})$, defined as
\begin{equation}\label{Eq:contform}
    \eta_{\mathcal{L}} = dz - \theta_{\mathcal{L}}
\end{equation}
We refer to the triple $(T^{2k-1}Q\times\mathbb{R},\eta_{\mathcal{L}},\mathcal{L})$, endowed with these dynamical elements, as a \textit{$k^{\textrm{th}}$ order (pre -) contact Lagrangian system}, where the qualitative `pre' is applicable when $\omega_{\mathcal{L}}$ is a closed, but degenerate 2 - form.\\

Since $T^{2k-1}Q\times\mathbb{R}$ is a standard contact manifold, whose treatment is encompassed within the usual contact framework \cite{bravetti2017contact,de2019contact,de2020review,etnyre2001introductory}, we may immediately state that the contact Lagrangian problem is to find a vector field $X_{\mathcal{L}}\in \mathfrak{X}^{\infty}(T^{2k-1}Q\times\mathbb{R})$ which satisfies
\begin{equation}\label{Eq:contactEL}
    i_{\scriptscriptstyle X_{\mathcal{L}}} d\eta_{\mathcal{L}} = dE_{\mathcal{L}} - \mathcal{R}_{\mathcal{L}}(E_{\mathcal{L}})\eta_{\mathcal{L}}\quad\quad\quad\quad i_{\scriptscriptstyle X_{\mathcal{L}}} \eta_{\mathcal{L}} = -\, E_{\mathcal{L}}
\end{equation}
Here, $\mathcal{R}_{\mathcal{L}}$ is the Reeb vector field for $T^{2k-1}Q\times \mathbb{R}$, defined uniquely via the equations
\begin{equation}\label{Eq:reeb}
    i_{\scriptscriptstyle \mathcal{R}_{\mathcal{L}}} d\eta_{\mathcal{L}} = 0\quad\quad\quad\quad i_{\scriptscriptstyle \mathcal{R}_{\mathcal{L}}} \eta_{\mathcal{L}} = 1
\end{equation}
These geometrical equations may equivalently be expressed in terms of the canonical prolongation of curves to the space $T^{2k-1}Q\times \mathbb{R}$. In particular, the physical trajectories of the system are curves $\phi_{\mathcal{L}}:\mathbb{R}\rightarrow Q\times\mathbb{R}$, which, in local coordinates, we write as $\phi_{\mathcal{L}}(t) = (q^i(t),z(t))$. The canonical prolongation of $\phi_{\mathcal{L}}$ to $T^{2k-1}Q\times\mathbb{R}$, which we denote via $\tilde{\phi}_{\mathcal{L}}$, has local coordinate expression
\begin{equation*}
    \tilde{\phi}_{\mathcal{L}}(t) = \left(q^i(t),\frac{dq^i}{dt}\biggr|_t,\,\,\cdots\,\,,\frac{d^{2k-1}q^i}{dt^{2k-1}}\biggr|_t,z(t) \right)
\end{equation*}
The geometrical equations (\ref{Eq:contactEL}) are then equivalent to
\begin{equation}\label{Eq:contactEL2}
    \begin{split}
        \sum_{\alpha=0}^{k}(-1)^{\alpha} D^{\alpha}_{\mathcal{L}}\left(\frac{\partial \mathcal{L}}{\partial q^i_{\alpha}}\right)\biggr|_{\tilde{\phi}_{\mathcal{L}}} =\, 0\\
        \frac{dz}{dt} = \mathcal{L}
    \end{split}
\end{equation}
\subsection{The Legendre - Ostrogradsky Map}\label{Subsec:LegContact}
Continuing to draw parallels with the symplectic case, the Legendre - Ostrogradsky map is a bundle morphism $\mathcal{FL}:T^{2k-1}Q\times \mathbb{R}\rightarrow T^*(T^{k-1}Q)\times \mathbb{R}$, which acts on the natural coordinates $(q^i_{\alpha},\widehat{p}^{\;\alpha}_i, z)$ of $T^{2k-1}Q\times \mathbb{R}$ in the following way
\begin{equation*}
    \mathcal{FL}^*\, q^i_{\alpha} = q^i_{\alpha} \quad\quad \mathcal{FL}^*\, p_i^{\alpha} = \widehat{p}_i^{\;\alpha}  \quad\quad \mathcal{FL}^* \, z = z
\end{equation*}
As before, the dynamical elements $\theta_{\mathcal{L}}$, $\omega_{\mathcal{L}}$, and $\eta_{\mathcal{L}}$ are related to their corresponding objects on $T^*(T^{k-1}Q)\times \mathbb{R}$ according to
\begin{equation*}
    \mathcal{FL}^*\,\theta=\theta_{\mathcal{L}}\quad\quad \mathcal{FL}^*\,\omega=\omega_{\mathcal{L}} \quad\quad \mathcal{FL}^*\,\eta=\eta_{\mathcal{L}}
\end{equation*}
\subsection{Hamiltonian Formalism}\label{Subsec:HamContact}
If $\mathcal{L}$ is a regular Lagrangian, we may introduce the unique Hamiltonian function $\mathcal{H}^c\in C^{\infty}(T^*(T^{k-1}Q)\times \mathbb{R})$, which satisfies $\mathcal{FL}^*\mathcal{H}^c=E_{\mathcal{L}}$. We then declare $(T^*(T^{k-1}Q)\times \mathbb{R},\eta,\mathcal{H}^c)$ to be a \textit{$k^{\textrm{th}}$ order contact Hamiltonian system}, and seek a vector field $X_{ \mathcal{H}}\in\mathfrak{X}^{\infty}(T^*(T^{k-1}Q)\times\mathbb{R})$ such that
\begin{equation}\label{Eq:contactHam}
    i_{\scriptscriptstyle X_{\mathcal{H}}} d\eta=d\mathcal{H}^c - \mathcal{R}(\mathcal{H}^c)\eta\quad\quad\quad\quad i_{\scriptscriptstyle X_{\mathcal{H}}} \eta = - \,\mathcal{H}^c
\end{equation}
in which $\mathcal{R}$ denotes the Reeb vector field on $T^*(T^{k-1}Q)\times \mathbb{R}$.\\

Suppose that an integral curve $\psi_{\mathcal{H}}: \mathbb{R}\rightarrow T^*(T^{k-1}Q)\times\mathbb{R}$ of $X_{\mathcal{H}}$ is expressed in local coordinates as $\psi_{\mathcal{H}}(t) = (q^i_{\alpha}(t),p^{\alpha}_i(t),z(t))$; the $k^{\textrm{th}}$ order Hamilton equations then become
\begin{equation}\label{Eq:contactHamlocal}
    \begin{split}
        \dot{q}^i_{\alpha}= \frac{\partial \mathcal{H}^c}{\partial p^{\alpha}_i}\biggr|_{\scriptscriptstyle \psi_{\mathcal{H}}(t)}&\quad\quad\quad \dot{p}^{\alpha}_i =-\left(\frac{\partial \mathcal{H}^c}{\partial q^i_{\alpha}} + p_i^{\alpha}\,\frac{\partial \mathcal{H}^c}{\partial z} \,\right) \biggr|_{\scriptscriptstyle \psi_{\mathcal{H}}(t)}\\
        &\dot{z} = \left( p^{\alpha}_i \,\frac{\partial \mathcal{H}^c}{\partial p^{\alpha}_i} - \mathcal{H}^c\right)\biggr|_{\scriptscriptstyle\psi_{\mathcal{H}}(t)}
    \end{split}
\end{equation}
Finally, in order to conclude our introduction to higher - order geometrical mechanics, we shall return to the example of the modified Pais - Uhlenbeck oscillator, demonstrating how this may be analysed within the contact framework we have developed.
\subsection{The Pais - Uhlenbeck Oscillator Revisited}\label{Subsec:PURevisited}
Fundamentally, our motivation for introducing contact geometry, is that the standard symplectic framework is ill - suited to describe physical systems which exhibit non - conservative effects \cite{ciaglia2018contact,gryb2021scale}. This is made particularly evident, when examining the Liouville theorem for symplectic manifolds, which asserts that the phase space volume occupied by solutions is conserved under the Hamiltonian flow \cite{koszul2019introduction,burkard2014classical}. The analogous statement for contact manifolds does \textit{not} hold. Instead, we find that phase space volumes undergo focusing and spreading under the Hamiltonian flow \cite{sloan2021scale}, and it is precisely this feature which makes contact manifolds the ideal arena in which to study non - conservative systems.\\

In light of this, we may illustrate the details of higher - order contact geometry introduced above, by adding a simple linear damping term to the Pais - Uhlenbeck Lagrangian (\ref{Eq:PUL})
\begin{equation}\label{Eq:dampedPU}
    \mathcal{L} = \frac{1}{2}\left(\dot{q}^2 - q^2\dot{\theta}^2 - \lambda\ddot{q}^2\right) - \gamma z
\end{equation}
in which $\gamma$ is some positive real constant, parameterising the damping of the oscillator. We find the Jacobi - Ostrogradsky momenta, just as we did for the free oscillator; however, we are now to use the Lagrangian total derivative $D_{\mathcal{L}}$, instead of $d_T$. Consequently, we find that
\begin{equation*}
    \widehat{p}_q^{\;1} = \frac{\partial \mathcal{L}}{\partial \ddot{q}} = -\,\lambda\ddot{q} \quad\quad\quad\quad \widehat{p}_q^{\;0} = \frac{\partial \mathcal{L}}{\partial \dot{q}} - D_{\mathcal{L}}(\widehat{p}_{q}^{\;1}) = \dot{q}+\lambda\dddot{q} - \lambda\gamma \ddot{q} \quad\quad\quad\quad \widehat{p}_{\theta}^{\; 0} = \frac{\partial\mathcal{L}}{\partial \dot{\theta}} = -\, q^2\dot{\theta}
\end{equation*}
Additionally, the energy function is identical in structure\footnote{We emphasise that it is \textit{structurally} similar - the momentum $\widehat{p}_q^{\;0}$ is clearly different to that of the free oscillator.} to that of the free oscillator, except for the presence of the damping term $\gamma z$
\begin{equation*}
    E_{\mathcal{L}} = \dot{q}\,\widehat{p}_q^{\;0} - \frac{1}{2}\left(\dot{q}^2 + \frac{1}{q^2}(\widehat{p}^{\;0}_{\theta})^2 + \frac{1}{\lambda}(\widehat{p}_q^{\;1})^2\right) + \gamma z
\end{equation*}
Suppose we express the Hamiltonian vector field $X_{\mathcal{H}}$, corresponding to $\mathcal{H}^c$, in local coordinates as
\begin{equation*}
    X_{\mathcal{H}} = A_0\frac{\partial}{\partial q} + A_1 \frac{\partial}{\partial \dot{q}} + A_2 \frac{\partial}{\partial \theta} + B_0\frac{\partial}{\partial p_q^0} + B_1\frac{\partial}{\partial p_q^1}+ B_2\frac{\partial}{\partial p_{\theta}^0} + C\frac{\partial}{\partial z}
\end{equation*}
The coordinate - free expressions (\ref{Eq:contactHam}) then become
\begin{align*}
    A_0 &= \frac{\partial \mathcal{H}^c}{\partial p_q^0} = \dot{q} & B_0 &= -\left(\frac{\partial \mathcal{H}^c}{\partial q} + p_q^0\frac{\partial \mathcal{H}^c}{\partial z}\right) = -\,\left(\frac{1}{q^3}(p^0_{\theta})^2 + \gamma p^0_q \right) \\
    A_1&= \frac{\partial \mathcal{H}^c}{\partial p_q^1}=-\,\frac{1}{\lambda}p_q^1 & B_1 &=  -\left(\frac{\partial \mathcal{H}^c}{\partial \dot{q}} + p_q^1\frac{\partial \mathcal{H}^c}{\partial z}\right) = \dot{q} - p_q^0 - \gamma p_q^1\\
    A_2 &= \frac{\partial \mathcal{H}^c}{\partial p^0_{\theta}}=-\,\frac{1}{q^2}p^0_{\theta} & B_2 & = -\left(\frac{\partial \mathcal{H}^c}{\partial \theta} + p^0_{\theta}\frac{\partial \mathcal{H}^c}{\partial z}\right) = -\, \gamma p^0_{\theta}
\end{align*}
\vspace{-6mm}
\begin{equation*}
    C = p_q^0\frac{\partial \mathcal{H}^c}{\partial p_q^0} + p_q^1\frac{\partial \mathcal{H}^c}{\partial p_q^1} + p_{\theta}^0\frac{\partial \mathcal{H}^c}{\partial p_{\theta}^0} - \mathcal{H}^c =\frac{1}{2}\left(\dot{q}^2 - \frac{1}{q^2}(p^0_{\theta})^2 - \frac{1}{\lambda}(p^1_q)^2\right) - \gamma z
\end{equation*}
Having developed all of the mathematical tools of higher - order geometry, and shown how this somewhat abstract formalism leads to a very natural way of describing Lagrangian and Hamiltonian mechanics, we may now introduce the concept of scaling symmetries, which play a central role in the main focus of this article: contact reduction. 

\section{Scaling Symmetries and Dynamical Similarity}\label{Sec:ScalingSym}

In general, a vector field $Y\in \mathfrak{X}^{\infty}(M)$ on a manifold $M$, of dimension $n$, is said to constitute a \textit{dynamical similarity} of $X\in \mathfrak{X}^{\infty}(M)$ if $[Y,X]= f X$, for some (generally) non - constant function $f:M\rightarrow \mathbb{R}$.\\

Given a (first - order) Hamiltonian system $(T^*Q,\omega,\mathcal{H})$, we refer to a vector field $D\in \mathfrak{X}^{\infty}(T^*Q)$ as a \textit{scaling symmetry} of degree $\Lambda$ if it satisfies:
\begin{itemize}
    \item $\mathfrak{L}_D\,\omega=\omega$
    \item $\mathfrak{L}_D\mathcal{H}=\Lambda \mathcal{H}$
\end{itemize}
in which $\mathfrak{L}$ denotes the Lie derivative. A short calculation reveals that
\begin{equation}\label{Eq:DynamicalSim}
    [D,X_{\mathcal{H}}] = (\Lambda-1)\,X_{\mathcal{H}}
\end{equation}
and so, we conclude that scaling symmetries are a particular class of dynamical similarity, for which the `function' defined above is in fact constant.\\

The fundamental interest of scaling symmetries lies in the fact that they map indistinguishable physical solutions into each other \cite{bravetti2017contact,sloan2018dynamical,sloan2021scale}. More specifically, our equations of motion are derived from an action principle, constructed as an integral of the 1 - form $\mathcal{L}\,dt$. Under the action of a scaling symmetry, this object undergoes an overall scaling by some constant factor $\lambda$. In order to carry out the extremisation of the action, we restrict variations to lie along all unscaled directions; it is then clear that those curves which produce stationary values the action, such that $\delta S=0$, will also satisfy $\delta (\lambda S) = 0$. Thus, solutions to the equations of motion (for the unscaled coordinates) remain solutions of the transformed system.\\

Recall that, as outlined in the introduction, our objective is to provide a framework which minimally encompasses the phenomena we wish to describe. Following Leibniz's Principle of the Identity of Indiscernibles (PII), we affirm that the action of a scaling symmetry, mapping one solution into a second, physically - indistinguishable solution, cannot possibly contribute to the closure of the algebra of dynamical observables \cite{rodriguez2014leibniz,frankel1981leibniz,brading2003symmetries}. Consequently, in line with our minimalist approach, this action should be excised from our ontology; this statement is made more mathematically - precise below.\\

Finally, we remark that, when seeking a quantum description of reality, it is precisely the observable degrees of freedom we shall wish to quantise \cite{huggett2001quantize}. As such, a theory formulated exclusively in terms of these observables is highly attractive.

\section{First - Order Contact Reduction}\label{Sec:FirstOrderContRed}

\subsection{Lagrangian Approach}\label{Subsec:LagApproach}
The generalisation of contact reduction to theories of higher order is most readily understood within the Lagrangian framework. As such, we shall dedicate this section to analysing the first - order case, motivating each step, so that the generalisation should seem relatively natural.\\

In general, we say that a symplectic Lagrangian system $(TQ,\omega_{\mathcal{L}},\mathcal{L})$ admits a scaling symmetry of degree $\Lambda$ if there exists a vector field $D\in \mathfrak{X}^{\infty}(TQ)$, such that
\begin{itemize}
    \item $\mathfrak{L}_D\,\theta_{\mathcal{L}} = \theta_{\mathcal{L}}$
    \item $\mathfrak{L}_D\mathcal{L}=\Lambda \mathcal{L}$
\end{itemize}
Additionally, we say that a function $x:TQ\rightarrow\mathbb{R}$ is a \textit{scaling function} of $D$ if it satisfies $\mathfrak{L}_Dx=x$.\\

We may imagine that the flow of $D$ traces out a curve within the space of solutions, and that physical measurements are impervious to displacements along this curve. Two points on the same integral curve are to be identified, and we use the symbolic notation $TQ/D$ to denote the reduced space, formed by the quotient under this identification.\\

In passing to the space $TQ/D$, we are excising from our description the unobservable notion of overall scale. It thus follows that $\textrm{dim}\,TQ/D = \textrm{dim}\,TQ - 1$, and so the reduced space is a contact manifold, upon which the dynamics of the original system are reproduced, without reference to the global scale, which was present within the symplectic description. Note that, since the degree of freedom we have eliminated is not a measurable quantity, we are able to provide a physically - equivalent description, specifying one datum fewer.\\ 

In our treatment of contact mechanics, we considered Lagrangian functions that were dependent upon the coordinates $q^i$, $\dot{q}^i$, and also a real parameter $z$. It was Herglotz who first considered that we might identify this parameter with the action itself \cite{herglotz1930beruhrungstransformationen,schillebeeckx1970gesammelte}, thereby obtaining a Lagrangian of the form $\mathcal{L}^H(q^i,\dot{q}^i,S)$.\\

The Herglotz - Lagrange equations are derived from a variational principle, in which we extremise the action
\begin{equation}\label{Eq:Herglotzvar}
    S = \int_{t_1}^{t_2} dt\; \mathcal{L}^H(q^i,\dot{q}^i,S)
\end{equation}
subject to the condition $\dot{S}=\mathcal{L}^H$ \cite{massa2023herglotz}. Carrying out this variational calculation,\footnote{See appendix (\ref{Appendix:A}) for details of how this (and the corresponding higher - order generalisation) is done.} we find that the equations of motion are given by
\begin{equation}\label{Eq:Herglotzeqns}
    \frac{d}{dt}\left(\frac{\partial \mathcal{L}^H}{\partial \dot{q}^i}\right)-\frac{\partial \mathcal{L}^H}{\partial q^i} = \frac{\partial \mathcal{L}^H}{\partial S}\frac{\partial \mathcal{L}^H}{\partial \dot{q}^i}
\end{equation}
In light of this, consider introducing a new parameter $\rho$, such that 
\begin{equation*}
    \dot{\rho}=-\,\frac{\partial \mathcal{L}^H}{\partial S}
\end{equation*}    
A short calculation then shows that the symplectic Lagrangian
\begin{equation}\label{Eq:fulllagrangian}
    \mathcal{L} = e^{\rho}(\mathcal{L}^H+\dot{\rho} S)
\end{equation}    
has equations of motion which coincide with those of $\mathcal{L}^H$, when restricted to the contact manifold, upon which $\mathcal{L}^H$ is defined.\\

Motivated by this, let $\mathcal{L}\in C^{\infty}(TQ)$ be a Lagrangian admitting a scaling symmetry $D$ of degree $\Lambda$. Without loss of generality, suppose that $D$ acts to scale a particular configuration space variable $Q$, whilst leaving $q$ unchanged. We look to replace the variable $Q$ with a scaling function $x(Q)$, and adopt coordinates along this direction. To do so, we make the time reparameterisation $d\tau = x^{\Lambda-1}\,dt$, so that
\begin{equation}\label{Eq:simplifiedvect}
    D=x\frac{\partial}{\partial x} + x'\frac{\partial}{\partial x'}
\end{equation}
in which primes are to be interpreted as derivatives with respect to $\tau$. Identifying $x=e^{\rho}$ we may then write our Lagrangian in the form
\begin{equation}\label{Eq:rhoLag}
    \mathcal{L}=e^{\Lambda\rho}f(\rho',q,q')
\end{equation}
In our new coordinates $(\rho,q,\rho',q')$, the scaling symmetry vector field is simply $D=\partial_{\rho}$.\\

Note that the Lagrangian (\ref{Eq:rhoLag}) has a factor of $e^{\Lambda\rho}$, and not simply $e^{\rho}$, as required by (\ref{Eq:fulllagrangian}). However, since the equations of motion are derived from an action principle, it is the Lagrangian form $\mathcal{L}\,dt$ which must reproduce invariant dynamics. Consequently, we construct
\begin{equation}\label{Eq:Lhat}
    \mathcal{L}\,dt \equiv \widehat{\mathcal{L}}\,d\tau \quad\implies \quad \widehat{\mathcal{L}} = e^{\rho} f(\rho',q,q')
\end{equation}
In the case that the scaling symmetry is already of the form (\ref{Eq:fulllagrangian}), so that no reparameterisations need to be made, we must proceed slightly differently. The validity of our formalism is predicated on the assumption that the Lagrangian is expressed as $e^{\rho} f(\dot{\rho},q,\dot{q})$. Clearly, if $d\tau=dt$, then our procedure will yield a final Lagrangian of the form $e^{\Lambda\rho} f(\dot{\rho},q,\dot{q})$. Consequently, we should instead make the identification $x = e^{\,\rho/\Lambda}$, with all subsequent steps implemented without modification.\\

In order to identify the action and Herglotz Lagrangian, we use the decomposition (\ref{Eq:Lhat}) of $\widehat{\mathcal{L}}$, and manipulate the following Euler - Lagrange equation for $\rho$
\begin{equation}\label{Eq:ELeqnsRho}
    \frac{d}{d\tau}\biggr(\frac{\partial \widehat{\mathcal{L}}}{\partial \rho'}\biggr) - \frac{\partial \widehat{\mathcal{L}}}{\partial \rho} = 0
\end{equation}
into the more suggestive form
\begin{equation}\label{Eq:f}
    f(\rho',q,q') = \frac{d}{d\tau} \left(\frac{\partial f}{\partial \rho'}\right) + \rho'\frac{\partial f}{\partial \rho'}
\end{equation}
Recalling that $\mathcal{L}^H=S'$, it is clear that we should identify
\begin{equation}
    S = \frac{\partial f}{\partial \rho'}\quad\quad\quad \mathcal{L}^H = f - \rho'S
\end{equation}
Note that (\ref{Eq:ELeqnsRho}) does \textit{not} yield the same equations of motion as the original system: the introduction of the time reparameterisation affects (amongst other things) the notion of energy within the transformed system. Thus, it is \textit{not} true to affirm that the Herglotz Lagrangian obtained by means of this reduction process has identical equations of motion to those of the original Lagrangian. We are instead constructing a simpler system, whose space of solutions intersects that of the full system on a surface of constant energy. At present, we content ourselves with acknowledging this fact, and postpone the explanation of how we recover the underlying system until section (\ref{Sec:Couplings}), where we discuss at length the implications of introducing a new time parameterisation.\\

As things stand, it would seem that we have done a great deal of work, and gained very little, and so it is worth reiterating why this reduction process is advantageous. Algebraically speaking, it is quite clear that we have passed to an equally, if not more complex description of our original system. However, in doing so, we have managed to capture the underlying dynamics, referencing only those quantities which are physically observable. Thus, we have exchanged superficial mathematical simplicity for a more fundamental description of the underlying physics.
\subsection{Hamiltonian Approach}\label{Subsec:HamApproach}
Let us now summarise the analogous reduction process on the Hamiltonian side. Suppose that $(T^*Q,\omega,\mathcal{H})$ is a symplectic Hamiltonian system, admitting a scaling symmetry $G$ of degree $\Lambda$, where we use $G$, so as to distinguish it from the Lagrangian $D$.\\

The space $T^*Q/G$ is, once again, a contact manifold of dimension $\textrm{dim}\,T^*Q-1$. In order to carry out the reduction process, we first identify a scaling function $x:T^*Q\rightarrow \mathbb{R}$, and construct a contact form and Hamiltonian on $T^*Q/G$, according to
\begin{equation}\label{Eq:reduction}
    \pi^*\eta=\frac{i_G\,\omega}{x}\quad\quad\quad\quad \pi^*\mathcal{H}^c=\frac{\mathcal{H}}{x^{\,\Lambda}}
\end{equation}
where $\pi:T^*Q\rightarrow T^*Q/G$ denotes the submersion between $T^*Q$ and the reduced space. The presence of the scaling function introduces a time reparamterisation $t\rightarrow\tau$, with $d\tau = x^{\Lambda-1}dt$. Consequently, an integral curve $\psi(\tau) = (q^i(\tau),p_i(\tau),S(\tau))$ of the Hamiltonian vector field corresponding to $\mathcal{H}^c$ satisfies
\begin{equation}\label{Eq:reducedeqns}
    \begin{split}
        \frac{dq^i}{d\tau} = \frac{\partial \mathcal{H}^c}{\partial p_i}\biggr|_{\scriptscriptstyle \psi(\tau)} & \quad\quad\quad\quad \frac{dp_i}{d\tau} = - \,\biggr(\frac{\partial \mathcal{H}^c}{\partial q^i}+ p_i\frac{\partial \mathcal{H}^c}{\partial S}\biggr)\biggr|_{\scriptscriptstyle \psi(\tau)}\\
        & \frac{dS}{d\tau}= \,\biggr( p_i \frac{\partial \mathcal{H}^c}{\partial p_i} - \Lambda \mathcal{H}^c\biggr)\biggr|_{\scriptscriptstyle \psi(\tau)}
    \end{split}
\end{equation}
Examples of the process of contact reduction within the Hamiltonian setting may be found in \cite{sloan2021scale,sloan2018dynamical,bravetti2023scaling,sloan2021new}, for example.

\section{Higher - Order Contact Reduction}\label{Sec:HigherOrderContRed}

The steps given in section (\ref{Subsec:LagApproach}) may easily be generalised to higher - order theories. However, on the space $TQ$, the conditions $\mathfrak{L}_D\,\theta_{\mathcal{L}} = \theta_{\mathcal{L}}$ and $\mathfrak{L}_D\mathcal{L} =\Lambda\mathcal{L}$ were sufficient to restrict $D$ in such a way that the change of coordinates to $x$ and $\tau$ always left a vector field of the form (\ref{Eq:simplifiedvect}). In order to guarantee the same for the higher - order case, we shall impose the additional constraint that our scaling symmetry be the tangent lift of a vector field $\mathcal{D}\in \mathfrak{X}^{\infty}(Q\times\mathbb{R})$, defined on the extended configuration space.\\

Without loss of generality, we may always choose coordinates on $Q\times \mathbb{R}$, such that $\mathcal{D}$ takes the form
\begin{equation}\label{Eq:generalvectfield}
    \mathcal{D} = AQ\frac{\partial}{\partial Q} + Bt\frac{\partial}{\partial t}
\end{equation}
in which $Q$ is our configuration space variable to be scaled, leaving all other $q^i$ unaffected, and $A$ and $B$ are real coefficients, parameterising the scaling \cite{sloan2023herglotz}.\\

Having identified a symmetry on the extended configuration space, we now lift this to a vector field $D$ on $T^{2k-1}Q$, in a manner which reproduces the scaling effects of $\mathcal{D}$. It is not difficult to see that the vector field
\begin{equation}\label{Eq:lifted}
    D = AQ\frac{\partial}{\partial Q} + \sum_{\alpha=1}^{2k-1} \left( (A-\alpha B) \, Q_{\alpha}\frac{\partial}{\partial Q_{\alpha}} - \alpha B\,q^i_{\alpha}\frac{\partial}{\partial q^i_{\alpha}} \right)
\end{equation}
has the desired effect. Evidently, $x = Q^{1/A}$ is a scaling function of $D$, and in order to adopt coordinates along this direction, we make the time reparameterisation
\begin{equation}\label{Eq:reparam}
    d\tau = x^{-B}dt = x^{\Lambda-1}dt
\end{equation}
where, in the second equality, we have used the fact that $B+\Lambda=1$. Consequently, our vector field $D$ now takes the form
\begin{equation}\label{Eq:simpD}
    D = \sum_{\alpha=\,0}^{2k-1}x^{(\alpha)}\frac{\partial}{\partial x^{(\alpha)}}
\end{equation}
in which $x^{(\alpha)}$ represents the $\alpha^{\textrm{th}}$ derivative of $x$ with respect to $\tau$. Defining $x=e^{\rho}$, and constructing $\widehat{\mathcal{L}}$, just as in the first - order case, we arrive at the Lagrangian
\begin{equation}\label{Eq:higherorderL}
    \widehat{\mathcal{L}}=e^{\rho}f(\rho',\rho'',\,\cdots, q^i_{\alpha'})
\end{equation}
where $q^i_{\alpha'}$ are all the unscaled coordinates and their corresponding $\tau$ derivatives, up to order $k$. We now manipulate the $k^{\textrm{th}}$ order Euler - Lagrange equation for $\rho$
\begin{equation}\label{Eq:ELforRho}
    \sum_{\alpha=0}^k(-1)^{\alpha}\frac{d^{\alpha}}{d\tau^{\alpha}} \biggr(\frac{\partial \widehat{\mathcal{L}}}{\partial \rho^{(\alpha)}}\biggr) = 0
\end{equation}
into the following schematic form
\begin{equation*}
    f - \rho'\,\Box = \frac{d}{d\tau}\,\Box 
\end{equation*}
From this, we may immediately identify the quantity $\Box$ as our action $S$, and a short calculation reveals that
\begin{equation}\label{Eq:generalisedS}
    S = \sum_{\alpha=0}^{k-1}(-1)^{\alpha}\left(\rho' + \frac{d}{d\tau}\right)^{\alpha} \frac{\partial f}{\partial \rho^{(\alpha+1)}}
\end{equation}
A small amount of algebraic manipulation reveals that the action is related to the Jacobi - Ostrogradsky momentum $\widehat{p}^{\;0}_{\rho}$ according to $S=e^{-\rho}\,\widehat{p}^{\;0}_{\rho}$. This fact will be of greater significance when discussing the corresponding Hamiltonian reduction process; however, it is noteworthy that we observe a generalisation of a feature of the first - order case. For theories defined on $T^*Q$, we always find that the action $S$ is (up to an overall multiplicative factor) equal to the momentum conjugate to the scaling variable.\\

In order to complete the reduction process, we introduce the variable $\chi$, with a new set of indices $a$, $b$, $c$, ... such that  $\chi^{(a)} = \rho^{(\alpha+1)}$. Writing $\mathcal{L}^H=f-\chi S$, we obtain a Herglotz Lagrangian that is of order $k-1$ in the variable $\chi$.\\

In complete analogy to the first - order case, if the scaling symmetry vector field is already of the form (\ref{Eq:simpD}), so that no time reparameterisation is required, we should instead identify $x=e^{\,\rho/\Lambda}$, and proceed as above.\\

In the above discussion, we have focused exclusively on the scaling variable $\rho$, and have neglected to demonstrate that our various reparameterisations do not somehow interfere with the dynamics of the remaining unscaled coordinates. We should, therefore, illustrate that the Euler - Lagrange equations for the $q^i$ are reproduced by the corresponding Herglotz Lagrangian. In order to do so, we consider the expression for the original Lagrangian\footnote{Here, for the sake of clarity, we have omitted the various pullbacks of $S$ and $\mathcal{L}^H$ to the reduced space $T^{k}Q/D$.}, written in terms of $\mathcal{L}^H$ and $S$
\begin{equation}\label{Eq:OriginalInvertedLagrangian}
    \widehat{\mathcal{L}} = e^{\rho}(\,\mathcal{L}^H + \rho'S)
\end{equation}
The Euler - Lagrange equations for the unscaled coordinates $q^i_{\alpha}$ read
\begin{align*}
    0 = \sum_{\alpha=0}^k(-1)^{\alpha}\frac{d^{\alpha}}{d\tau^{\alpha}} \biggr(\frac{\partial \widehat{\mathcal{L}}}{\partial q^i_{\alpha}}\biggr) = \sum_{\alpha=0}^k(-1)^{\alpha}\left(e^{\rho}\frac{d^{\alpha}}{d\tau^{\alpha}}\frac{\partial\mathcal{L}^H}{\partial q^i_{\alpha}} + \left(\frac{d^{\alpha}}{d\tau^{\alpha}}e^{\rho}\right)\frac{\partial\mathcal{L}^H}{\partial q^i_{\alpha}}\right)
\end{align*}
Recalling that the quantity $\rho$ was introduced in such a way that $\rho' = -\,\frac{\partial \mathcal{L}^H}{\partial S}$, we see that the terms on the right hand side of the above equation may be rewritten as follows
\begin{equation*}
    \sum_{\alpha=0}^k(-1)^{\alpha}\left(e^{\rho}\frac{d^{\alpha}}{d\tau^{\alpha}}\frac{\partial\mathcal{L}^H}{\partial q^i_{\alpha}} + \left(\frac{d^{\alpha}}{d\tau^{\alpha}}e^{\rho}\right)\frac{\partial\mathcal{L}^H}{\partial q^i_{\alpha}}\right) = \sum_{\alpha=0}^k (-1)^{\alpha}\,e^{\rho} \left(\frac{d}{d\tau} - \frac{\partial \mathcal{L}^H}{\partial S}\right)^{\alpha}\frac{\partial \mathcal{L}^H}{\partial q^i_{\alpha}}
\end{equation*}
Finally, from our construction of the Herglotz Lagrangian $\mathcal{L}^H=f-\rho'S$, it is clear that $\frac{\partial\mathcal{L}^H}{\partial q^i_{\alpha}}$ is independent of $S$; consequently, we may freely add a term of the form $\mathcal{L}^H\frac{\partial}{\partial S}$ to the bracket on the right hand side of the above, to yield
\begin{equation*}
    0= \sum_{\alpha=0}^k (-1)^{\alpha}\,e^{\rho} \biggr(\underbrace{\frac{d}{d\tau} + \mathcal{L}^H\frac{\partial}{\partial S} -\frac{\partial \mathcal{L}^H}{\partial S}}_{\stackrel{(\ref{Eq:Lagtot})}{=}\;D_{\mathcal{L}}}\,\biggr)^{\alpha}\,\frac{\partial \mathcal{L}^H}{\partial q^i_{\alpha}} = \sum_{\alpha=0}^k (-1)^{\alpha}\,e^{\rho}\,D_{\mathcal{L}}^{\alpha}\left(\frac{\partial\mathcal{L}^H}{\partial q^i_{\alpha}}\right)
\end{equation*}
We thus see that the Euler - Lagrange equation for the unscaled coordinates leads to the corresponding Herglotz equation; inverting this argument, it is apparent that the Herglotz equations for the $q^i$ correctly reproduce the original unscaled dynamics, when restricted to the contact manifold $T^kQ/D$.\\

Finally, we should inspect the Herglotz - Lagrange equation for the reduced coordinate $\chi$. Since $\chi = -\,\frac{\partial \mathcal{L}^H}{\partial S}$, and the function $f(\chi,\chi',\cdots,q^i_{\alpha})$ is independent of $\rho$, it follows that the Lagrangian total derivative may be expressed as $D_\mathcal{L} = (d_T+\chi)$. Consequently, we have
\begin{equation}
    0 = \sum_{a=0}^{k-1}(-1)^{a} D^{a}_{\mathcal{L}}\left(\frac{\partial \mathcal{L}^H}{\partial \chi^{(a)}}\right) = -\,S + \sum_{a=0}^{k-1}(-1)^a(d_T+\chi)^a \left(\frac{\partial f}{\partial \chi^{(a)}} \right)
\end{equation}
which is simply a rearrangement of the definition (\ref{Eq:generalisedS}) of $S$. The above discussion highlights an important point: since we have constructed our formalism directly from the Euler - Lagrange equation (\ref{Eq:ELforRho}), the Herglotz - Lagrange equation (\ref{Eq:contactEL2}) for the reduced coordinate is trivially satisfied. Thus, the original equation of motion for $\rho$ is reproduced within the contact system by imposing that $S'= f - \chi S$.\\

Topologically speaking, this reduction process is quite different to that of the first - order case: for a Lagrangian on $TQ$, we use a vector field, defined over the same space, to completely eliminate the scaling coordinate, and its derivatives. In this way, the resulting reduced space is isomorphic to $T\widehat{Q}\times\mathbb{R}$, in which $\widehat{Q}$ denotes a configuration space manifold composed of all unscaled coordinates. In contrast, in the higher - order case, we use a vector field defined on $T^{2k-1}Q$ to eliminate reference to the scaling coordinate $\rho$, within the Lagrangian, which is a function on $T^kQ$. For this reason, we retain all derivatives of $\rho$, and our contact manifold instead has the topology $T^kQ/\,\mathbb{R}$.

\section{Hamiltonian Formulation}\label{Sec:HamFormulationHigherOrder}

Having demonstrated how the process of contact reduction may be extended to theories of higher order within the Lagrangian formalism, we would like to see how this translates over to the Hamiltonian setting. We shall continue to work with the Lagrangian $\widehat{\mathcal{L}}$, expressed in terms of $\rho$, whose corresponding energy function is given by
\begin{equation}\label{Eq:rhoenergyfunc}
    E_{\mathcal{L}} = \sum_{\alpha=0}^{k-1} \left(\rho^{(\alpha+1)} \widehat{p}^{\;\alpha}_{\rho} + q^i_{\alpha+1}\widehat{p}^{\;\alpha}_i\right) - \widehat{\mathcal{L}}
\end{equation}
Performing a Legendre transform on $E_{\mathcal{L}}$, we express the Hamiltonian $\mathcal{H}$ in terms of $(\rho^{(\alpha)},p^{\alpha}_{\rho},q^i_{\alpha},p^{\alpha}_i)$. In these coordinates, the 2 - form $\omega\in\Omega^2(T^*(T^{k-1}Q))$ is expressed as
\begin{equation}\label{Eq:2form}
    \omega = \sum_{\alpha=0}^{k-1}\left(d\rho^{(\alpha)} \wedge\, dp^{\alpha}_{\rho} + dq^i_{\alpha}\wedge dp^{\alpha}_i \right)
\end{equation}
We then find that the vector field
\begin{equation}\label{Eq:vecfieldG}
    G = \frac{\partial}{\partial \rho} + \sum_{\alpha=0}^{k-1} \left(p^{\alpha}_{\rho} \frac{\partial}{\partial p^{\alpha}_{\rho}} + p^{\alpha}_i\frac{\partial}{\partial p^{\alpha}_i}\right)\;\;\in\;\;\mathfrak{X}^{\infty}(T^*(T^{k-1}Q))
\end{equation}
is a degree - one scaling symmetry. Introducing the submersion $\beta:T^*(T^{k-1}Q)\rightarrow T^*(T^{k-1}Q)/G$, the contact form $\eta$ is found from
\begin{equation}\label{Eq:reducedcontactform}
    \beta^*\eta = \frac{i_G\,\omega}{e^{\rho}} = e^{-\rho}\left[ dp^0_{\rho} - p^0_{\rho}\,d\rho -\sum_{\alpha=1}^{k-1} p^{\alpha}_{\rho}\,d\rho^{(\alpha)} - \sum_{\alpha=0}^{k-1} p^{\alpha}_i\,dq^i_{\alpha} \right]
\end{equation}
The first two terms may be combined into the single derivative $d(e^{-\rho}p_{\rho}^0)$, which, as we have seen, coincides precisely with the action $S$. Consequently, we complete the reduction process by defining the following variables on $T^*(T^{k-1}Q)/G$
\begin{align}\label{Eq:redvbls}
        \beta^*\chi^{(a)} &= \rho^{(\alpha+1)}  &  \beta^*q^i_{\alpha} &= q^i_{\alpha} & \beta^*S= \frac{p^0_{\rho}}{e^{\rho}} \notag \\
        \beta^*\pi^{a}_{\chi} &= \frac{p^{\alpha+1}_{\rho}}{e^{\rho}} & \beta^*\pi^{\alpha}_i &= \frac{p^{\alpha}_i}{e^{\rho}} & &
\end{align}
As in the Lagrangian case, all $\rho$ coordinates have been redefined in terms of $\chi$, and now carry an index $a$, which assumes values from $0$ to $k-2$. It then follows that the contact form may be written as
\begin{equation}\label{Eq:Hcontactform}
    \eta = dS -\sum_{a=0}^{k-2}\pi^{a}_{\chi}\,d\chi^{(a)} - \sum_{\alpha=0}^{k-1} \pi^{\alpha}_i\,dq^i_{\alpha}
\end{equation}
while the contact Hamiltonian $\mathcal{H}^c$ is, as usual, given by
\begin{equation}\label{Eq:contactHamiltonian}
    \beta^*\mathcal{H}^c=\frac{\mathcal{H}}{e^{\rho}}
\end{equation}
Coordinates on $T^*(T^{k-1}Q)/G$ are $(\chi^{(a)},\pi^a_{\chi},q^i_{\alpha},\pi^{\alpha}_i,S)$, and, as we shall see when working with explicit examples, the equation of motion for the coordinate $S$ 
\begin{equation}\label{Eq:EOMforS}
   \frac{dS}{d\tau}= \pi^a_{\chi}\frac{\partial \mathcal{H}^c}{\partial \pi^a_{\chi}} + \pi^{\alpha}_i\frac{\partial \mathcal{H}^c}{\partial \pi^{\alpha}_i} - \mathcal{H}^c
\end{equation}
gives precisely the Herglotz Lagrangian $\mathcal{L}^H$, upon substituting the momenta for their Legendre - transformed counterparts. Additionally, as for the Lagrangian case, we should demonstrate that the equations of motion for the unscaled coordinates $q^i$ are correctly reproduced within our symmetry - reduced framework. The dynamical equation for the momentum $\pi^{\alpha}_i$ is (dropping the projection map $\beta$)
\begin{align*}
    \frac{d}{d\tau}\pi^{\alpha}_i &= -\,\left(\frac{\partial\mathcal{H}^c}{\partial q^i_{\alpha}} + \pi^{\alpha}_i\frac{\partial \mathcal{H}^c}{\partial S}\right)\\
    \frac{d}{d\tau}(e^{-\rho}p^{\alpha}_i) &= -\,\left(e^{-\rho}\frac{\partial\mathcal{H}}{\partial q^i_{\alpha}} + e^{-\rho}\rho'p^{\alpha}_i\right)\\
    \implies\quad\frac{d}{d\tau} p^{\alpha}_i &= -\,\frac{\partial\mathcal{H}}{\partial q^i_{\alpha}}
\end{align*}
where, in passing from the first to the second line, we have used the fact that if $\frac{\partial \mathcal{L}^H}{\partial S}=-\,\rho'$, then $\frac{\partial\mathcal{H}^c}{\partial S}=\rho'$. Consequently, we see that our formalism correctly reflects the dynamics of the unscaled variables in both the Lagrangian and Hamiltonian settings.\\

The contact Hamiltonian may alternatively be obtained by performing the reduction within the Lagrangian formalism, before carrying out a Legendre transform between the reduced spaces. In this case, we take coordinates on $T^{2k-1}Q/D$ to be $(\chi^{(a)},\widehat{\Pi}^a_{\chi},q^i_{\alpha},\widehat{\Pi}^{\alpha}_i,S)$, where
\begin{equation}\label{Eq:Pimomenta}
    \begin{split}
        \widehat{\Pi}^a_{\chi} = \sum_{r=0}^{k-a-2} (-1)^{r} D_{\mathcal{L}}^{r}\left(\frac{\partial \mathcal{L}^H}{\partial \chi^{(a+r+1)}}\right)\\
        \widehat{\Pi}^{\alpha}_{i} = \sum_{r=0}^{k-\alpha-1} (-1)^{r} d_T^{r}\left(\frac{\partial \mathcal{L}^H}{\partial q^i_{\alpha+r+1}}\right)
    \end{split}
\end{equation}
Note that the momenta of the unscaled coordinates are calculated using $d_T$, as opposed to $D_{\mathcal{L}}$. Upon taking the Legendre transform of $\mathcal{L}^H$, we obtain a contact Hamiltonian, expressed in terms of the variables $(\chi^{(a)},\Pi^a_{\chi},q^i_{\alpha},\Pi^{\alpha}_i,S)$. We then find that the equations of motion for the $\chi^{(a)}$ and their conjugate momenta $\Pi^a_{\chi}$ exactly reproduce the recursion relation (\ref{Eq:rec}) of the symplectic momenta, upon making the identification $\chi=\rho'$ and $p_{\rho}^{\alpha+1}= e^\rho \Pi^a_{\chi}$. Further, the equation of motion for $\Pi^0_{\chi}$ is given by
\begin{equation}
    \frac{d}{d\tau} \Pi_\chi^0 = \frac{\partial f}{\partial \chi} - S - \Pi_\chi^0 \chi
\end{equation}
which, following similar steps to those given above, may be rearranged to reproduce the expression $e^\rho S = p_\rho^0$.

\section{The Evolution of Couplings}\label{Sec:Couplings}

So far, we have considered the (Lagrangian) scaling symmetry to act only on the space $T^{2k-1}Q$. However, recall that the effect of these transformations is to rescale the units used to measure physical quantities, without changing the overall dynamics. In the Kepler problem, for example, we shall see that physical solutions are impervious to the scaling $(r,t)\rightarrow (\lambda^2 r,\lambda^3 t)$, which effectively redefines what we mean by metres and seconds. Any physical theory often depends upon a number of empirically - determined coupling constants, and so we should consider the possibility that our scaling symmetry may also act on these parameters. In this case, we say that the couplings become \textit{dynamical} in nature. The manner in which we handle dynamical couplings in the first - order case is well - established, particularly within the Hamiltonian setting \cite{bravetti2023scaling}, and so we shall simply outline the procedure in what follows.
\subsection{The First - Order Case: An Overview}\label{Subsec:FirstOrderCouplings}
Strictly speaking, our motivation for considering dynamical couplings is twofold: on the one hand, it is often possible to identify a scaling symmetry of a given system, even when its uncoupled analogue admits no such transformation. As an example, we might consider the following Lagrangian
\begin{equation*}
    \mathcal{L} = \frac{1}{2}\dot{r}^2 + \frac{1}{2}r^2\dot{\theta}^2 + \frac{1}{r} + r^2
\end{equation*}
representing a simple, uncoupled Kepler system, with an additional $r^2$ term. The presence of this new term destroys the symmetry of $\mathcal{L}$ under the transformation $(r,t)\rightarrow (\lambda^2 r,\lambda^3 t)$, and there is no obvious combination of rescaling that restores this symmetry. However, were we to introduce a pair of coupling constants $C,D\in\mathbb{R}$, and rewrite our Lagrangian as
\begin{equation*}
    \mathcal{L} = \frac{1}{2}\dot{r}^2 + \frac{1}{2}r^2\dot{\theta}^2 + \frac{C}{r} + Dr^2
\end{equation*}
then we see that $(r,D,t)\rightarrow (\lambda^2 r,\lambda^{-6}D,\lambda^3 t)$ restores the overall scaling $\mathcal{L}\rightarrow \lambda^{-2}\mathcal{L}$.\\

In order to allow the scaling symmetry to act on the coupling parameters of our theory we promote these `constants' to velocities with no corresponding conjugate position variables\footnote{By this, we mean that they exist, but $\mathcal{L}$ has no dependence on them, and so they act as dummy variables.}. In this way, the Euler - Lagrange equations immediately tell us that these velocities are constant in time. Supposing that our Lagrangian depends upon $N$ (positive) coupling parameters, this construction amounts to extending the tangent bundle $TQ$ to include two copies of $\mathbb{R}^N$: one for the set of new velocities, and another for their corresponding dummy position variables.\\

As a result, the scaling symmetry now acts on the extended manifold $TQ\times \mathbb{R}^{N}\times\mathbb{R}^N_+$, and we may carry out the reduction process exactly as in section (\ref{Subsec:LagApproach}). In the complementary Hamiltonian picture, we apply a similar procedure, promoting the couplings instead to \textit{momenta}, with dummy configuration space variables. Hamilton's equations then guarantee that these momenta will be constant in time.\\

It may seem counterintuitive to introduce variables in place of constants: our goal, after all, is to \textit{reduce} the amount of information required to describe the evolution of a system. However, in order to fully integrate the original system, we would need to specify each of the initial positions and velocities, as well as provide numerical values for the couplings. In the reduced system, by contrast, each of the coupling parameters has been replaced by a constant velocity, and so we have effectively exchanged constants for boundary conditions. Since the reduction process also eliminates the notion of overall scale, we are in fact justified in claiming that we require fewer data to evolve the observables of our system in time.
\subsection{Time Reparameterisation}\label{Subsec:TimeReparam}
The second reason for which we are particularly interested in promoting coupling constants to dynamical variables, is that, as we alluded to in section (\ref{Subsec:LagApproach}), any system upon which we impose a new time parameterisation ceases to correctly reproduce the equations of motion of the original Lagrangian.\\

This is hardly surprising, given that the transformation from $t$ to $\tau$ introduces a spatial dependence, through the factor of $e^{(\Lambda-1)\rho}$. In the reparameterised system, the energy does not transform in the same way as in the original Lagrangian, and it is this feature which creates the discrepancy between the systems' equations of motion. However, the solutions obtained under the new time parameterisation are in fact valid, provided we are working with systems of zero total energy. This fact is of particular relevance in the context of General Relativity; one of our main goals in presenting an extension of the symmetry reduction process to higher - order theories, is to explore the evolution of physically - motivated, higher - order cosmological models, in a setting in which overall scale is not present. General Relativity, being a time - reparameterisation - invariant theory, necessarily has a vanishing Hamiltonian, and so falls within the class of zero - energy systems mentioned above.\\

For those theories which do not enjoy time reparameterisation invariance, configurations of non - zero energy require additional steps to recover the original dynamics. In particular, we add an overall constant $E>0$ to the original Lagrangian, representing the total energy of the system. This constant then becomes dynamical in nature within the reparameterised system. The most illustrative way to appreciate this concept is through example; let us therefore consider adding an overall constant $E$ to the Lagrangian of the standard Kepler system:
\begin{equation}\label{Eq:NonzeroEKepler}
    \mathcal{L} = \frac{1}{2}\dot{r}^2 + \frac{1}{2}r^2\dot{\theta}^2 + \frac{1}{r} \pm E   
\end{equation}
where the $\pm$ allows for the possibility of either open or closed orbits \cite{sloan2023herglotz}. Following the steps outlined in section (\ref{Subsec:LagApproach}), we see that, identifying $e^{\rho} = r^{1/2}$, and introducing a new time parameterisation $d\tau = e^{-3\rho}dt$, the Lagrangian $\widehat{\mathcal{L}}$, as defined in (\ref{Eq:Lhat}), becomes
\begin{equation}\label{Eq:LhatKepler}
    \widehat{\mathcal{L}} = e^{\rho}\left(2\rho^{\prime\,2} + \frac{1}{2}\theta^{\prime\,2} + 1\right) \pm Ee^{3\rho}
\end{equation}
Note that, what was a constant energy in one time parameterisation, has now assumed the role of a potential term in the transformed system. This is precisely why solutions of zero energy pose no problems: multiplication of zero by some $\rho$ - dependent factor will always yield zero. In order to reproduce the original system's dynamics, $E$ must become dynamical. Thus, let us promote $E\rightarrow \dot{z}^{\lambda}$, where the constant power $\lambda$ is determined, imposing that the reparameterised energy transform appropriately. It is not difficult to see that with $\lambda=2/3$, we have
\begin{equation}\label{Eq:CorrectedLhatE}
    \widehat{\mathcal{L}} = e^{\rho}\left(2\rho^{\prime\,2} + \frac{1}{2}\theta^{\prime\,2} + 1\pm z^{\prime\,2/3}\right)
\end{equation}
The key conclusion we draw from this analysis, is that a constant energy may always be traded for a potential term, by making a transformation to a new time parameterisation. In the Hamiltonian picture, our system evolves on the constraint surface $\mathcal{H}=E$; however, in light of the previous comment, we know that there is always a time reparameterisation that we can make, in which the new Hamiltonian has precisely zero energy, and so, in practice, we consider the evolution of $\mathcal{H}' = \mathcal{H}-E$ on the constraint surface $\mathcal{H}'=0$.
\subsection{Higher - Order Generalisation}\label{Subsec:HigherOrderCouplings}
The procedure outlined above, in which we promote constants to velocities (or momenta, in the Hamiltonian case) carries over in precisely the same way to theories of higher order. Since the original (Lagrangian) scaling symmetry was a vector field over the space $T^{2k-1}Q$, we promote the $N$ (positive) coupling constants to velocity variables $\dot{z}^i$ such that $\mathcal{L}$ has no dependence upon the $z^i$, or any of their derivatives of order $\geqslant 2$. In this way, we allow the scaling symmetry to act on the extended tangent bundle $T^{2k-1}Q\times \mathbb{R}^N\times \mathbb{R}^N_+$, and proceed to reduce the system exactly as in section (\ref{Sec:HigherOrderContRed}).\\

The manner in which we treat systems of non - zero energy is also identical to that of the first - order case: we promote the constant representing the total energy to a velocity variable $E\rightarrow \dot{z}^{\lambda}$, and determine the power $\lambda$ by requiring that the energy transform appropriately in the new time parameterisation.

\section{Examples}\label{Sec:Examples}

Having presented our results on generalised contact reduction, we shall now demonstrate how this process is carried out in practice, via three examples. Since we have used the Pais - Uhlenbeck oscillator as a reference guide throughout, to illustrate the framework of higher - order geometrical mechanics, we shall complete the narrative, and construct the symmetry - reduced system, as our first example.\\

The aims of this section are both practical and didactic; as such, our second example is chosen to highlight all the features of the reduction process, but has no apparent physical motivation. The final example, by contrast, is of great physical relevance, but is far simpler, and thus less illustrative, from a practical perspective. Given the relevance of this third example, we shall also discuss the utility of the symmetry - reduced system, from both a calculational, and physical point of view.
\subsection{The Pais - Uhlenbeck Oscillator}\label{Eq:ReductionofPUOscillator}
The majority of the work necessary to construct the symmetry - reduced Pais - Uhlenbeck oscillator has been carried over the course of the paper; here we bring these elements together, in order to illustrate the reduction procedure. Recall that the Lagrangian is given by
\begin{equation}\label{Eq:PaisLagrangian}
    \mathcal{L} = \frac{1}{2}\left(\dot{q}^2 - q^2\dot{\theta}^2 - \lambda\ddot{q}^2\right)
\end{equation}
It is clear that under the transformation $(q,\theta,t) \rightarrow (kq,\theta,t)$, the Lagrangian is rescaled by a multiplicative factor of $k^2$; lifting this to a vector field on $T^3Q$, we obtain
\begin{equation}
    D= q \frac{\partial}{\partial q} + \dot{q} \frac{\partial}{\partial \dot{q}} + \ddot{q} \frac{\partial}{\partial \ddot{q}} + \dddot{q} \frac{\partial}{\partial \dddot{q}}
\end{equation}
From the Lagrangian vector field (\ref{Eq:PUXLinq}), we see that $[D,X_\mathcal{L}]=1$; additionally, it is clear from (\ref{Eq:PUOmegaLinq}) that $\mathfrak{L}_D \omega_\mathcal{L} = d \circ i_{D} \omega_\mathcal{L} = \omega_\mathcal{L}$ and $\mathfrak{L}_D \mathcal{L} = 2 \mathcal{L}$. Both of these observations demonstrate that $D$ is a scaling symmetry of degree two.\\

The base manifold is $Q \simeq \mathbb{R} \times S^1$; however, conservation of angular momentum ensures that the configuration space separates into two dynamically - disjoint pieces, consisting of positive and negative values of $q$. Thus, without loss of generality, we restrict to positive values and, in accordance with our general procedure, define $q=e^{\,\rho/2}$, with the result that
\begin{equation}\label{Eq:PaisRhoLagrangian}
    \mathcal{L}= \frac{1}{2}e^{\rho} \left(\frac{1}{4}\dot{\rho}^2 - \dot{\theta}^2 - \frac{\lambda}{16}\left(4\dot{\rho}^2 + 4\dot{\rho}^2\ddot{\rho} + \dot{\rho}^4 \right) \right)
\end{equation}
and the scaling symmetry is simply $D=\partial_{\rho}$. The Jacobi - Ostrogradsky momenta are then
\begin{equation*}
    \widehat{p}^{\;1}_{\rho} = \frac{\partial\mathcal{L}}{\partial \ddot{\rho}} = -\, \frac{\lambda e^\rho}{4}\left(\ddot{\rho}+\frac{\dot{\rho}^2}{2} \right) \quad\quad\quad \widehat{p}^{\;0}_{\rho} = \frac{\partial\mathcal{L}}{\partial \dot{\rho}} - d_T(\widehat{p}^{\;1}_{\rho}) = \frac{e^\rho}{4} \left( \dot{\rho} + \lambda \dot{\rho}\ddot{\rho} + \lambda \dddot{\rho} \right) \quad  \quad\quad \widehat{p}^{\;0}_{\theta} = \frac{\partial\mathcal{L}}{\partial \dot{\theta}}= -\, e^\rho \dot{\theta}
\end{equation*}
whilst the equations of motion read
\begin{equation*}
    \rho^{(4)} +2\dddot{\rho}\ddot{\rho}+\frac{3}{2}\ddot{\rho}^2+\frac{3}{2}\ddot{\rho}\dot{\rho}^2+\frac{1}{8}\dot{\rho}^4 + \frac{1}{\lambda}\left( \frac{1}{2}\dot{\rho}^2 +\ddot{\rho} +2\dot{\theta}^2 \right) =0 \quad\quad\;\textrm{and}\quad\quad\; \ddot{\theta}+\dot{\rho}{\dot{\theta}}=0
\end{equation*}
Note that, as expected, only derivatives of $\rho$ (and not $\rho$ itself) appear in these equations of motion. Application of (\ref{Eq:generalisedS}) to the Lagrangian (\ref{Eq:PaisRhoLagrangian}) confirms that $S=e^{-\rho} \widehat{p}^{\;0}_{\rho}$, and the reduction is completed introducing $\chi=\dot{\rho}$, and constructing the Herglotz Lagrangian
\begin{equation}\label{Eq:PUHergLag}
    \mathcal{L}^H = \frac{1}{2}\left(\frac{1}{4}\chi^2 - \dot{\theta}^2 - \frac{\lambda}{16}\left(4\chi^2 + 4\chi^2\dot{\chi} + \chi^4 \right) \right) - \chi S
\end{equation}
Let us suppose, for pedagogical reasons, that we had simply been given equation (\ref{Eq:PUHergLag}) as our starting point; then, under the identification $\chi=\dot{\rho}$, the equation of motion for $\chi$ recovers the following expected result
\begin{equation*}
    \frac{d}{dt}\left(\frac{\partial\mathcal{L}^H}{\partial \dot{\chi}}\right) - \frac{\partial\mathcal{L}^H}{\partial \chi} = \frac{\partial\mathcal{L}^H}{\partial \dot{\chi}}\frac{\partial\mathcal{L}^H}{\partial S} \quad\quad\;\implies\quad\quad\; S = \frac{1}{4}\chi +\frac{\lambda}{4}\chi \dot{\chi}^2 +\frac{\lambda}{4} \ddot{\chi} = e^{-\rho}\widehat{p}^{\;0}_{\rho}
\end{equation*}
Similarly, the equation of motion for $\theta$
\begin{equation*}
      \quad \ddot{\theta}+\chi \dot{\theta}=0
\end{equation*}
exactly coincides with the result obtained above, from the symplectic Lagrangian. Finally, the equation of motion of the coordinate $\rho$ is obtained setting $\dot{S}=\mathcal{L}^H$.\\

The Hamiltonian, expressed in the variables $\rho$ and $\theta$, is given by
\begin{equation}\label{Eq:PaisRhoHamiltonian}
     \mathcal{H} = \dot{\rho}p^0_{\rho} -\frac{2 (p_{\rho}^1)^2}{\lambda e^\rho}-\frac{1}{2} \dot{\rho}^2 p_{\rho}^1 - \frac{e^\rho}{8}\dot{\rho}^2 - \frac{(p_{\theta}^0)^2}{2e^\rho}
\end{equation}
with corresponding symplectic form 
\begin{equation}
    \omega = d\rho \wedge dp_{\rho}^0+ d\dot{\rho}\wedge dp_{\rho}^1 + d\theta \wedge dp^0_\theta
\end{equation}
It is then a simple exercise to verify that the vector field
\begin{equation*}
    G = \frac{\partial}{\partial\rho} + p^0_{\rho}\frac{\partial}{\partial p^0_{\rho}} + p^1_{\rho}\frac{\partial}{\partial p^1_{\rho}} + p^0_{\theta}\frac{\partial}{\partial p^0_{\theta}}
\end{equation*}
does indeed constitute a scaling symmetry of degree one. The contact Hamiltonian may now be obtained, either via a direct Legendre transform of the Herglotz Lagrangian (\ref{Eq:PUHergLag}), or following the reduction scheme presented in the previous section. In both cases, we arrive at 
\begin{equation}\label{Eq:PaisContactHamiltonian}
    \mathcal{H}^c =  - \left( \frac{1}{2}(\pi_{\theta})^2 + \frac{1}{2}\chi^2 \pi_{\chi} + \frac{2}{\lambda}(\pi_{\chi})^2 + \frac{1}{8}\chi^2\right) + \chi S
\end{equation}
where we have omitted the momentum superscripts, as there is only one of each type. The contact form for the reduced space is given by
\begin{equation}
     \eta = dS - \pi_{\chi} d\chi - \pi_{\theta}d\theta
\end{equation}
Finally, from the contact Hamiltonian (\ref{Eq:PaisContactHamiltonian}), we find the following equations of motion
\begin{align}
    \dot{\chi} &= -\frac{4}{\lambda} \pi_{\chi} - \frac{1}{2}\chi^2 &  \dot{\pi}_{\chi} &= - 2\chi \pi_{\chi}  + \frac{1}{4}\chi -S \notag \\
    \dot{\theta}&=-\pi_\theta & \dot{\pi}_{\theta} &= -\chi \pi_\theta
\end{align}
\vspace{-10mm}
\begin{equation*}
    \dot{S}= \frac{1}{8}\chi^2 -\frac{2}{\lambda}(\pi_{\chi})^2-\frac{1}{2} (\pi_\theta)^2- \chi S
\end{equation*}
As expected, these may be rearranged and combined to reproduce the definition of $S$ in terms of the unscaled variables. Note also that the final equation is simply $\dot{S} = \mathcal{L}^H$, expressed in momentum variables.
\subsection{Modified Kepler System}\label{Subsec:EgKepler}
Our next example is that of a modified Kepler system; the planar Kepler problem is an extremely well - studied example in classical physics \cite{brouwer2013methods,chobotov2002orbital}, which we have already used several times throughout. Working in the centre of mass frame, temporarily setting all masses and couplings to unity, the Lagrangian is given by
\begin{equation}\label{Eq:keplerL}
    \mathcal{L} = \frac{1}{2}\dot{r}^2 + \frac{1}{2}r^2\dot{\theta}^2 + \frac{1}{r} + r^{3/4}\dddot{r}^{1/2}
\end{equation}
The base manifold over which we work is $Q\simeq \mathbb{R}^+\times S^1$, parameterised via the usual polar coordinates $(r,\theta)$. Under the transformation $(r,t)\rightarrow (\lambda^2 r, \lambda^3 t)$, the Lagrangian scales as $\mathcal{L}\rightarrow \lambda^{-2}\mathcal{L}$, and so we may immediately write down the $(\Lambda=-\,2)$ scaling symmetry vector field
\begin{equation*}
    D = 2r\frac{\partial}{\partial r} - \dot{r}\frac{\partial}{\partial \dot{r}} -4 \ddot{r}\frac{\partial}{\partial \ddot{r}} -7 \dddot{r}\frac{\partial}{\partial \dddot{r}} -10 r^{(4)}\frac{\partial}{\partial r^{(4)}} -13 r^{(5)}\frac{\partial}{\partial r^{(5)}} -3\dot{\theta}\frac{\partial}{\partial \dot{\theta}}
\end{equation*}
where we have identified $Q= r$, $A=2$, and $B=3$ in (\ref{Eq:generalvectfield}).\\

Note that, unlike in the previous example, here, we must introduce a time reparameterisation. Indeed, the vector field $D$ is rendered of the form (\ref{Eq:simpD}) taking $x = r^{1/2}$, and $d\tau=x^{-3}dt$. Further identifying $x=e^{\rho}$, we obtain the following Lagrangian, written in terms of $\rho$ and $\tau$
\begin{equation}\label{Eq:KeplerLhat}
    \widehat{\mathcal{L}} = \mathcal{L}\,\frac{dt}{d\tau}=e^{\rho}\biggr(\underbrace{2\rho^{\prime\,2} + \frac{1}{2}\theta^{\prime\,2} + 1 + \sqrt{2\left(\rho''' - 6\rho'\rho'' + 4\rho^{\prime\,3}\right)}}_{\equiv\;f(\rho',\rho'',\rho''',\theta')}\,\biggr)
\end{equation}
The Herglotz Lagrangian is then
\begin{equation}\label{Eq:KeplerHerglotz}
    \mathcal{L}^H = 2\chi^2 + \frac{1}{2}\theta^{\prime\,2} + 1 + \sqrt{2\left(\chi'' - 6\chi\chi' + 4\chi^3\right)} - \chi S
\end{equation}
with $\chi = \rho'$; the equations of motion are deduced imposing the condition $S' = f-\chi S$. As discussed in section (\ref{Subsec:TimeReparam}), these solutions are only valid for systems of zero total energy. In order to obtain a more general description, applicable to systems of arbitrary energy $E$, we promote $E\rightarrow \dot{z}^k$, and see that with $k=2/3$, we obtain a reparameterised Lagrangian
\begin{equation}\label{Eq:KeplerLhatnonzeroE}
    \widehat{\mathcal{L}}=e^{\rho}\biggr(2\rho^{\prime\,2} + \frac{1}{2}\theta^{\prime\,2} + 1 + \sqrt{2\left(\rho''' - 6\rho'\rho'' + 4\rho^{\prime\,3}\right)}\pm z^{\prime\,2/3}\,\biggr)
\end{equation}
in which the energy transforms correctly. Note also that, had we chosen to include a coupling constant $C$ for the $1/r$ term, this would \textit{not} have become dynamical, with our choice of scaling symmetry.\\

Taking coordinates on $T^5Q/D$ to be $(\chi,\chi',\widehat{\Pi}^0_{\chi},\widehat{\Pi}^1_{\chi},\theta,\widehat{\Pi}^0_{\theta})$, after applying the Legendre - Ostrogradsky map to the energy function of $\mathcal{L}^H$, we obtain the following contact Hamiltonian
\begin{equation}\label{Eq:KeplerContactH1}
    \mathcal{H}^c = -\,2\chi^2  + \frac{1}{2}(\Pi^0_{\theta})^2 - 1 -\frac{1}{2\Pi^1_{\chi}} + \chi'\,\Pi^0_{\chi} + 6\chi\chi'\,\Pi^1_{\chi} -4\chi^3\,\Pi^1_{\chi} + \chi S
\end{equation}
The contact form $\eta$ is then given by
\begin{equation}\label{Eq:Keplercontactform1}
    \eta = dS - \Pi^0_{\chi}\,d\chi - \Pi^1_{\chi}\,d\chi' - \Pi^0_{\theta}\,d\theta
\end{equation}
As verification of the consistency of our formalism, note that the equation of motion for $S$ is
\begin{equation*}
    \frac{dS}{d\tau} = 2\chi^2 + \frac{1}{2}(\Pi_{\theta}^0)^2 + 1 + \frac{1}{\Pi^1_{\chi}} -\chi S
\end{equation*}
and substituting expressions for $\Pi^1_{\chi}$ and $\Pi^0_{\theta}$, in terms of $\chi$, $\theta$, and their derivatives, we recover precisely (\ref{Eq:KeplerHerglotz}), showing that $S' = \mathcal{L}^H$, as expected.\\

The Hamiltonian $\mathcal{H}\in C^{\infty}(T^*(T^2Q))$, corresponding to (\ref{Eq:KeplerLhat}), is given by
\begin{equation}\label{Eq:KeplerH}
    \mathcal{H} =-\,e^{\rho}\left(2\rho^{\prime\,2} + 1\right) + \frac{1}{2}e^{-\rho}(p^0_{\theta})^2 - \frac{e^{2\rho}}{2p^2_{\rho}} + \rho'' p^1_{\rho}+6\rho'\rho''p^2_{\rho} - 4\rho^{\prime\,3}p^2_{\rho} + \rho' p^0_{\rho}
\end{equation}
On the space $T^*(T^2Q)$, we have the symplectic form
\begin{equation}\label{Eq:Keplersympl}
    \omega = d\rho \wedge dp_{\rho}^0 + d\rho'\wedge dp_{\rho}^1 + d\rho''\wedge dp_{\rho}^2 + d\theta\wedge dp^0_{\theta}
\end{equation}
and we see that
\begin{equation}\label{Eq:KeplerscalingG}
    G = \frac{\partial}{\partial \rho} + p^0_{\rho}\frac{\partial}{\partial p^0_{\rho}} + p^1_{\rho}\frac{\partial}{\partial p^1_{\rho}} + p^2_{\rho}\frac{\partial}{\partial p^2_{\rho}} + p^0_{\theta}\frac{\partial}{\partial p^0_{\theta}}
\end{equation}
is a degree - one scaling symmetry. We use this to pass to the reduced space $T^*(T^2Q)/G$, upon which we define the coordinates
\begin{align}\label{Eq:Keplerreducedcoords}
        \beta^*\chi &= \rho' &\beta^*\chi' &= \rho'' &\beta^*\theta &= \theta\notag\\
        \beta^*\pi^0_{\chi} &= \frac{p^1_{\rho}}{e^{\rho}} &\beta^*\pi^1_{\chi} &= \frac{p^2_{\rho}}{e^{\rho}} &\beta^*\pi^0_{\theta} &= \frac{p^0_{\theta}}{e^{\rho}}
\end{align}
\vspace{-7mm}
\begin{equation*}    
    \beta^*S = \frac{p^0_{\rho}}{e^{\rho}}
\end{equation*}
Consequently, the contact form $\eta$, and Hamiltonian $\mathcal{H}^c$ may be written as
\begin{equation}\label{Eq:Keplercontact2}
    \begin{split}
        \eta &= dS- \pi^0_{\chi}\,d\chi - \pi^1_{\chi}\,d\chi' - \pi^0_{\theta}\,d\theta\\
        \mathcal{H}^c =-\,2\chi^2 + \frac{1}{2}&(\pi^0_{\theta})^2 - 1 - \frac{1}{2\pi^1_{\chi}} + \chi'\pi^0_{\chi}+6\chi\chi'\pi^1_{\chi} - 4\chi^3\pi^1_{\chi} + \chi S
    \end{split}
\end{equation}
coinciding perfectly with (\ref{Eq:Keplercontactform1}) and (\ref{Eq:KeplerContactH1}).\\

This modified Kepler system is of no (known) physical relevance, although it is not unreasonable to assume that structurally similar terms could arise from some kind of perturbative expansion. Nevertheless, we have demonstrated how it is possible to reduce the order of the Lagrangian, within the $\rho$ sector, obtaining a description cast only in terms of scale - reduced quantities. These quantities evolve as an autonomous subset of the original symplectic system, without the need to reference overall scale. Eliminating this degree of freedom, which is inaccessible to an intrinsic observer, offers a more fundamental description of the problem.
\subsection{Higher - Order FLRW System}\label{Subsec:EgFLRW}
Theories of modified gravity have been of increasing interest in recent years \cite{sotiriou2010f}; shortly after the publication of Einstein's paper in 1915, physicists such as Weyl and Eddington began to experiment with higher - order corrections to General Relativity, motivated principally by scientific curiosity \cite{weyl1919space,eddington1923mathematical}. However, increasing amounts of evidence from high - energy particle physics, observational astrophysics, and theoretical cosmology suggest that mere curiosity ought not be our sole motivation for considering such theories \cite{vilkovisky1992effective,starobinsky1980new}.\\

The formalism we have developed thus far is applicable only to higher - order theories of particles, and does \textit{not} extend to those theories whose dynamical objects are fields. Consequently, we may not consider General Relativity in its full field - theory formulation; instead, we shall restrict ourselves to the (flat) FLRW sector, which has potential physical relevance to our own universe \cite{green2014well}. We therefore consider a metric of the form
\begin{equation}\label{Eq:FLRWMetric}
    ds^2 =-\,N^2(t) \, dt^2 + a^2(t)\delta_{ij}\,dx^i dx^j
\end{equation}
in which $N(t)$ denotes the lapse function. One may easily verify that the non - zero connection coefficients for such a spacetime are
\begin{equation}\label{Eq:ConnectionCoeffs}
    \Gamma^0_{00} = \frac{\dot{N}}{N}\quad\quad\quad\quad \Gamma^0_{ij} = \frac{a\dot{a}}{N^2}\delta_{ij} \quad\quad\quad\quad \Gamma^i_{0j} = \Gamma^i_{j0} = \frac{\dot{a}}{a}\delta^i_j
\end{equation}
It then follows that the Ricci scalar is given by
\begin{equation}\label{Eq:Ricci}
    R = \frac{6}{N^2}\left[\frac{\ddot{a}}{a}\,+\,\biggr(\frac{\dot{a}}{a}\biggr)^2 - \frac{\dot{a}}{a}\frac{\dot{N}}{N}\right] = \frac{2}{N^2} \left[\frac{\ddot{v}}{v}- \frac{\dot{v}^2}{3v^2} - \frac{\dot{v}}{v}\frac{\dot{N}}{N} \right]
\end{equation}
where, in the second equality, we have chosen to introduce the volume scale factor $v=a^3$, for computational ease. Since our objective is to exemplify the symmetry reduction procedure, we shall henceforth choose to work in proper time $\tau$, setting the lapse function $N(t)$ to unity. Details of the calculation for arbitrary choice of non-trivial lapse are presented in appendix (\ref{Appendix:B}).\\

With the choice to work in proper time, we consider a modified Einstein - Hilbert action of the form
\begin{equation}\label{Eq:ModifiedEinsteinHilbert}
    S_{\scriptscriptstyle EH} = \frac{1}{2\kappa}\int_M d^4x \sqrt{-g} \;f(R) + \frac{1}{\kappa}\int_{\partial M} d^3y \sqrt{|h|} \;f'(R) K
\end{equation}
where we have supplemented the usual bulk action with an additional Gibbons - Hawking - York (GHY) surface term, suitably modified for the case of $f(R)$ theories \cite{Guarnizo_2010}. Here, $h$ is the determinant of the induced metric on the boundary $\partial M$, upon which we have local coordinates $y^{\alpha}$; $K$ refers to the trace of the extrinsic curvature of $\partial M$, and $f'(R) = df/dR$.\\

It is known that $f(R)$ models are dynamically equivalent to scalar - tensor theories \cite{teyssandier1983cauchy,whitt1984fourth,chiba20031}; additionally, within these higher - derivative theories, the variations of the first derivatives of the metric encode genuine dynamical degrees of freedom, and so should not remain arbitrary \cite{dyer2009boundary}. The constraint which must instead be imposed, is that variations of the Ricci scalar $R$ vanish at the boundary; with this, the modified GHY term in the action (\ref{Eq:ModifiedEinsteinHilbert}) correctly reproduces the system's dynamics, thereby preserving the equivalence with the scalar - tensor theory.\\

For this particular example, we shall choose the function $f(R) = R - \lambda R^2$, so that $f'(R) = 1-2\lambda R$. In addition to the purely gravitational sector, we also include a minimally - coupled, spatially - homogeneous scalar field $\varphi$, such that the complete action reads
\begin{equation}\label{Eq:CompleteAction}
    S = \frac{1}{2\kappa}\int_M d^4x \sqrt{-g} \;\left(R - \lambda R^2 + 2\kappa\left(\frac{1}{2}\ddot{\varphi}^2+ \frac{1}{2}\dot{\varphi}^2 - V(\varphi)\right)\right) + \frac{1}{\kappa}\int_{\partial M} d^3y\,\sqrt{|h|}\left(1-2\lambda R\right) K
\end{equation}
We have chosen to include a second - order term in the scalar field Lagrangian, so as to demonstrate that our formalism is still sound, even when the reduction variable is not the highest derivative term.\\

The spacetime manifold $M$ is globally hyperbolic, thus allowing a foliation into spacelike hypersurfaces $\Sigma$, whose comoving volume we denote $V_0$. The metric is both homogeneous and isotropic, and so the modified GHY term is proportional to $V_0f'(R)\dot{v}$, in which $f'(R)$ is to be replaced with the appropriate expression in terms of $v$ and $\dot{v}$. A complete analysis of the resulting surface terms, together with their effects on the final solution space is beyond the scope of this paper. Thus, we shall content ourselves with imposing the condition that $\dot{v}=0$ on the boundary, thereby eliminating the contribution from the GHY term.\\

In light of this, upon discarding total derivative terms, we obtain the following Lagrangian density\footnote{In order to avoid pedantic repetition of the words `Lagrangian density', we shall, in what follows, speak of the Lagrangian, and use the same symbol $\mathcal{L}$ we have employed throughout.}
\begin{equation}\label{Eq:FLRWlagrangian}
    \mathcal{L} = -\,\frac{1}{24\pi G}\frac{\dot{v}^2}{v}-\frac{\lambda}{4\pi G}\left(\frac{\ddot{v}^2}{v}-\frac{4\ddot{v}\dot{v}^2}{3v^2}+ \frac{5\dot{v}^4}{9v^3} \right) + v\left(\frac{1}{2}\ddot{\varphi}^2 +\frac{1}{2}\dot{\varphi}^2 - V(\varphi)\right)
\end{equation}
The configuration manifold $Q$ is parameterised via the two local coordinates $(v,\varphi)$, and we see that the transformation $(v,\varphi,t)\rightarrow (kv,\varphi,t)$ yields a corresponding change of $\mathcal{L}\rightarrow k \mathcal{L}$. Lifting this extended configuration space scaling symmetry to $T^3Q$, we obtain
\begin{equation}\label{Eq:FLRWscaling}
    D = v\frac{\partial}{\partial v} +\dot{v}\frac{\partial}{\partial\dot{v}} + \ddot{v}\frac{\partial}{\partial \ddot{v}} + \dddot{v}\frac{\partial}{\partial \dddot{v}}
\end{equation}
Since this vector field is already of the desired form, and is a scaling symmetry of degree one, we immediately identify $x = v = e^{\rho}$, and write the Lagrangian as 
\begin{equation}\label{Eq:FLRWfinalL}
    \mathcal{L} = -\, e^{\rho} \left[ \frac{1}{72\pi G} \left(3\dot{\rho}^2 + 2\lambda\left(2\dot{\rho}^4 + 6\dot{\rho}^2\ddot{\rho} + 9\ddot{\rho}^2\right)\right) - \frac{1}{2}\ddot{\varphi}^2-\frac{1}{2}\dot{\varphi}^2 +V(\varphi)\right]
\end{equation}
As in the previous example, we identify the function $f$, reduce the order within the $\rho$ sector, defining $\chi = \dot{\rho}$, and construct the Herglotz Lagrangian
\begin{equation}\label{Eq:FLRWHerglotz}
    \mathcal{L}^H =  -\, \frac{1}{72\pi G} \left(3\chi^2 + 2\lambda\left(2\chi^4 + 6\chi^2\dot{\chi} + 9\dot{\chi}^2\right) \right) + \frac{1}{2}\ddot{\varphi}^2 +\frac{1}{2}\dot{\varphi}^2- V(\varphi) - \chi S
\end{equation}
The advantages of our formalism are twofold: on the one hand, we have eliminated reference to the scale factor $a$ (through the volume $v$). It has been established that there is no impediment to the continuation of contact - reduced cosmological models through singularities, at which General Relativity ceases to provide well - defined solutions \cite{mercati2022traversing,hoffmann2024continuation}. It would be of great interest, therefore, to examine whether this continuation generalises to higher - order theories, in which we eliminate the scale factor, but retain its derivatives.\\

In addition to this, had we not used a scaling symmetry to eliminate $a$, and thus reduce the order of the theory to first derivatives in $\chi$, we would be obliged to treat this cosmological model within the metric $f(R)$ formulation \cite{li2007cosmology}. In this way, we avoid the need to introduce the equivalent scalar - tensor theory, as described above \cite{sotiriou20096,de2010f}, and may instead study the dynamics of this system using only those tools we have developed throughout. It should be emphasised, however, that we are, at present, limited to work within the cosmological sector, until we have a more complete formalism, which allows us to treat field theories of arbitrary order.\\

In order to perform a Legendre transform on the Herglotz Lagrangian (\ref{Eq:FLRWHerglotz}), we calculate
\begin{equation*}
    \widehat{\Pi}^{0}_{\chi} = \frac{\partial \mathcal{L}^H}{\partial \dot{\chi}} = -\,\frac{\lambda}{6\pi G}(\chi^2+3\dot{\chi})
\end{equation*}
\vspace{-5mm}
\begin{equation*}
        \widehat{\Pi}^1_{\varphi} = \frac{\partial \mathcal{L}^H}{\partial \ddot{\varphi}} = \ddot{\varphi}\quad\quad\quad\quad
        \widehat{\Pi}^0_{\varphi} = \frac{\partial \mathcal{L}^H}{\partial \dot{\varphi}} - d_T (\widehat{\Pi}^1_{\varphi})=\dot{\varphi} - \dddot{\varphi}
\end{equation*}
We then find that the contact Hamiltonian is given by
\begin{equation}\label{Eq:FLRWHC1}
        \mathcal{H}^c = \frac{1}{72\pi G}\left(3\chi^2+2\lambda\chi^4\right) -\frac{1}{3}\chi^2 \Pi^0_{\chi} -\frac{\pi G}{\lambda}(\Pi^0_{\chi})^2 + \dot{\varphi}\Pi^0_{\varphi} -\frac{1}{2}\dot{\varphi}^2+V(\varphi) +\frac{1}{2}(\Pi_{\varphi}^1)^2 + \chi S
\end{equation}
with corresponding (non - trivial) equations of motion
\begin{align}\label{Eq:FLRWEoM}
    \dot{\chi} &= -\,\frac{1}{3}\chi^2 - \frac{2\pi G}{\lambda} \Pi^0_{\chi} & \dot{\Pi}^0_{\chi} &= -\,\frac{1}{36\pi G}(3\chi + 4\lambda\chi^3) - \frac{1}{3}\chi\Pi^0_{\chi} - S \notag \\
    \dot{\Pi}^0_{\varphi} &=-\,\frac{\partial V}{\partial \varphi} - \chi\Pi^0_{\varphi} & \dot{\Pi}^1_{\varphi} &= -\Pi^0_{\varphi} + \dot{\varphi} - \chi\Pi^1_{\varphi} 
\end{align}
\vspace{-6mm}
\begin{equation*}
    \dot{S} = -\,\frac{1}{72\pi G}(3\chi^2+4\lambda\chi^4) -\frac{\pi G}{\lambda}(\Pi^0_{\chi})^2 + \frac{1}{2}\dot{\varphi}^2 - V(\varphi) +\frac{1}{2}(\Pi^1_{\varphi})^2 -\chi S
\end{equation*}
These equations highlight a particularly important aspect of symmetry - reduced cosmological models; specifically, the right hand side of the $\varphi$ momentum equations contain additional terms, which encode deviations from the expected behaviour of the scalar field in a flat background spacetime. These terms are similar in structure to those which arise from friction; this offers a novel insight into the character of cosmological expansion: when treated within a framework that dispenses with those structures which are not fundamentally required to close the algebra of the dynamical variables, the expansion of the universe is an emergent property, that manifests itself through friction - like terms, which alter the equations of motion of the observables.\\

It is now straightforward to perform a Legendre transform on the full Lagrangian (\ref{Eq:FLRWfinalL}), to obtain the Hamiltonian
\begin{equation}\label{Eq:FLRWfullH}
    \begin{split}
        \mathcal{H} = \frac{e^{\rho}}{72\pi G}\left(3\dot{\rho}^2+2\lambda\dot{\rho}^4\right)+ \dot{\rho} p^0_{\rho}-\frac{1}{3}\dot{\rho}^2 p^1_{\rho} -\frac{\pi G}{\lambda}e^{-\rho}(p^1_{\rho})^2 + \dot{\varphi}p^0_{\varphi}\\
        - \, e^{\rho}\left(\frac{1}{2}\dot{\varphi}^2-V(\varphi)\right) +\frac{1}{2}e^{-\rho}(p_{\varphi}^1)^2
    \end{split}
\end{equation}
As expected, we have the following scaling symmetry of degree one
\begin{equation}\label{Eq:FLRWG}
    G = \frac{\partial}{\partial \rho} + p^0_{\rho}\frac{\partial}{\partial p^0_{\rho}} + p^1_{\rho}\frac{\partial}{\partial p^1_{\rho}} + p^0_{\varphi}\frac{\partial}{\partial p^0_{\varphi}} + p^1_{\varphi}\frac{\partial}{\partial p^1_{\varphi}}
\end{equation}
The reduction process is then completed exactly as in the previous example: we take coordinates on the space $T^*(TQ)/G$ to be
\begin{align}\label{Eq:FLRWreducedcoords}
        \beta^*\chi &= \dot{\rho} &\beta^*\dot{\varphi} &= \dot{\varphi} &\beta^*\ddot{\varphi}& = \ddot{\varphi}\notag\\
        \beta^*\pi^0_{\chi} &= \frac{p^1_{\rho}}{e^{\rho}} &\beta^*\pi^0_{\varphi} &= \frac{p^0_{\varphi}}{e^{\rho}} &\beta^*\pi^1_{\varphi} &= \frac{p^1_{\varphi}}{e^{\rho}}
\end{align}
\vspace{-7mm}
\begin{equation*}
    \beta^*S = \frac{p^0_{\rho}}{e^{\rho}}
\end{equation*}
such that the contact form and Hamiltonian are given by
\begin{equation}\label{Eq:FLRWcontactform}
    \begin{split}
        &\eta = dS - \pi^0_{\chi}\,d\chi - \pi^0_{\varphi}\,d\varphi - \pi^1_{\varphi}\,d\dot{\varphi}\\
        \mathcal{H}^c = \frac{1}{72\pi G}\left(3\chi^2+2\lambda\chi^4\right) &-\frac{1}{3}\chi^2 \pi^0_{\chi} -\frac{\pi G}{\lambda}(\pi^0_{\chi})^2 + \dot{\varphi}\pi^0_{\varphi} - \frac{1}{2}\dot{\varphi}^2+V(\varphi)+\frac{1}{2}(\pi_{\varphi}^1)^2 + \chi S
    \end{split}
\end{equation}
Finally, since we made no adjustments to our parameterisation of time, we know that the contact equations of motion reproduce the dynamics of the original FLRW system, for solutions of arbitrary energy. In particular, we see that the condition $\dot{S} = f - \chi S$ yields
\begin{equation}\label{Eq:FLRWHerglotzEoM}
    \dddot{\chi} = \frac{1}{12\lambda}\left(\chi^2 + 2\dot{\chi}\right) - \chi^2\dot{\chi} - \frac{3}{2}\dot{\chi}^2 - 2\chi\ddot{\chi} +\frac{2\pi G}{\lambda}\left(\frac{1}{2}\ddot{\varphi}^2 + \frac{1}{2}\dot{\varphi}^2 - V(\varphi)\right)
\end{equation}
which, after a short calculation, substituting $\chi\rightarrow\dot{\rho}$, can be shown to coincide with the second - order Euler - Lagrange equations obtained directly from (\ref{Eq:FLRWfinalL}).

\section{Symplectification}\label{Sec:Symplectification}

Before concluding, it is important to highlight that, in all cases, we may recover our original theory by symplectifying the underlying contact system \cite{sloan2021new}. Since the contact form $\eta$ is defined only up to a multiplicative constant, we may consider introducing a real number $y$, defining the contact form $\tilde{\eta} = y\eta$. Working within the Hamiltonian formalism, we may then promote $y$ to a real variable, and declare $\omega = d(y\eta)$ to be a symplectic form on the space $(T^*(T^{k-1}Q)/G)\times \mathbb{R}$.\\

For example, taking the contact form in (\ref{Eq:FLRWcontactform}), we write $\omega=d(y\eta)$, and define momenta on $(T^*(TQ)/G)\times\mathbb{R}$ to be $P=y\pi$. Further identifying $P^{\,0}_{\chi}=S$, we write $\omega$ as
\begin{equation*}
    \omega = dy \wedge dP^{\,0}_{\chi} + d\chi\wedge dP^{\,1}_{\chi} + d\varphi\wedge dP^{\,0}_{\varphi} + d\dot{\varphi} \wedge dP^{1}_{\varphi}
\end{equation*}
In this way, we have embedded the underlying contact system, within a larger symplectic framework. The equations of motion derived from the Hamiltonian $\mathcal{H}=y\mathcal{H}^c$ are identical to those of $\mathcal{H}^c$; however, the system described by the symplectic Hamiltonian is now conservative in nature: the dissipative effects of $\mathcal{H}^c$ are compensated through the evolution of the variable $y$.\\

The crucial conclusion to be taken from this, is the following: we have argued throughout that a description of nature, from which all superfluous structure is excised, is truly more fundamental than one which predicts the same physical phenomena, but exhibits redundancies in its formulation. This line of reasoning led us to a framework in which the minimal set of elements of our theory were those degrees of freedom accessible via direct observation, and any (potentially unobservable) parameters required to evolve these observables in time, such that an autonomous set was obtained. It is thus highly unusual that, if physics is truly relational, we should be able to embed this non - conservative description within a symplectic framework, by introducing an overall immeasurable scale. Moreover, this symplectic description works extremely well for a broad class of systems. On the other hand, we may view this statement in another light: perhaps the embedding of a contact theory, based on relational terms, into a larger symplectic framework is simply a reflection of an inculcated tendency to view closed systems as fundamental, and thus more desirable.

\section{Conclusions and Outlook}\label{Sec:Conclusions}

The application of techniques of differential geometry to classical mechanics provides a highly elegant description of a multitude of physical systems, whilst also making manifest many of their interesting features. In this article, for example, we have seen a number of implications of the fact that, unlike their symplectic counterparts, contact manifolds are odd - dimensional. Most notable of these, were the non - conservation of the measure, and of mechanical energy, under the evolution of the system. It was precisely these features which made contact manifolds a well - suited arena in which to develop non - conservative frameworks. The need for such a formalism arose from a conceptual shift, in which we argued that precedence should be given to a minimally - sufficient description of physical law. Through use of the Principle of Essential and Sufficient Autonomy, together with Leibniz's Identity of Indiscernibles, we deduced that a relational description of nature necessarily inherited a non - conservative characteristic.\\

After presenting an introduction to the more formal geometrical elements needed to describe higher - order classical mechanics, we dedicated considerable effort to understanding the details of the symmetry reduction process for first - order theories. Building upon this, we observed that this procedure admitted a very natural generalisation to theories containing higher - order derivatives. Our primary motivation for considering higher - order theories lies in the desire to explore gravitational models, which go beyond the standard framework of General Relativity. As such, one of the examples used to illustrate the formalism we developed in the first part of the paper, was a simple $f(R)$ model, in a flat, homogeneous, and isotropic universe. Despite being highly idealised, this model still allowed us to appreciate certain features of the symmetry - reduced description, such as the manifestation of the expansion of the universe as a friction - like phenomenon.\\

The Pais - Uhlenbeck oscillator was explored as a running example, and provided a controlled environment, in which to explore the formalisms, and highlight the interesting mathematical features of our construction. This model required no rescaling of the time coordinate, and so the space of solutions to the symmetry - reduced system was directly comparable to that of the original symplectic configuration. To give a more complete example, in which our time parameterisation did require rescaling, we presented a modified Kepler system. This allowed us to illustrate the complications that arise, when temporal coordinates are rescaled. In particular, we observed that for configurations of non - zero energy, we were forced to introduce an additive constant to the Lagrangian, which then became dynamical in nature within the reparameterised system. The role of this new dynamical parameter was to ensure that the symmetry - reduced description continued to correctly reflect the dynamics of the original system, even when a new time parameterisation was introduced.\\

As outlined in section (\ref{Subsec:EgFLRW}), the development of a formalism applicable to field theories will have significant implications for our ability to treat more complex and realistic models of gravity. Some progress has been made in this area \cite{sloan2025dynamical}; however, the geometrical structures employed here are those of $k$ - symplectic and $k$ - contact manifolds \cite{de2015methods,gaset2020contact}. Anticipating that we shall seek to develop a framework adapted to treat field theories of arbitrary order, it will be of great interest to explore the process of first - order contact reduction, in the context of multisymplectic geometry \cite{prieto2014geometrical,roman2009multisymplectic} which, a priori, appears to be more amenable to higher - order generalisations.\\

An additional area of future investigation of particular interest, is that of quantisation. The field of geometric quantisation has received increasing attention in recent years, and the procedure for treating symplectic manifolds, and more generally, Poisson manifolds, is relatively well - established \cite{woodhouse1992geometric,vaisman1991geometric}. Unfortunately, the same is \textit{not} true of contact manifolds: the odd - dimensionality of these spaces greatly hinders our ability to canonically define a Poisson bracket structure - a prerequisite for any quantisation attempt. Some progress has been made in this area \cite{rajeev2008quantization,fitzpatrick2011geometric}; however, it is, as yet, unclear how to proceed.
\appendix

\section{Herglotz - Lagrange Equations: A Variational Derivation}\label{Appendix:A}

As mentioned in section (\ref{Subsec:LagApproach}), the Herglotz - Lagrange field equations may be derived from a variational approach, as opposed to via the geometric perspective we have presented throughout.\\

For simplicity, consider a Lagrangian $\mathcal{L}^H$, depending upon a single generalised coordinate $q$, and its time derivatives up to order $k$. We shall collectively denote these coordinates $q_{\alpha}$, with $0\leqslant \alpha\leqslant k$. In contrast to the familiar variational approach to Lagrangian mechanics, we also allow the function $\mathcal{L}^H$ to depend upon the action itself, so that we seek to extremise the action
\begin{equation}\label{Eq:HerglotzAction}
    S = \int_{t_1}^{t_2} dt\;\mathcal{L}^H(q_{\alpha},S)
\end{equation}
subject to the condition that $\dot{S}=\mathcal{L}^H$.\\

This condition is most readily enforced through use of a Lagrange multiplier $\lambda$, so that we vary the modified action
\begin{equation}\label{Eq:ModifiedAction}
    \mathcal{Z}(q_{\alpha},S,\dot{S},t) = \int_{t_1}^{t_2} dt\;\left(\mathcal{L}^H + \lambda(\mathcal{L}^H-\dot{S})\right)
\end{equation}
subject only to the usual constraints on admissable variations. We parameterise such variations according to
\begin{equation}\label{Eq:Variations}
     \begin{split}   
        q(t) \,\longrightarrow\, q(t) + \varepsilon \eta(t)\\
        S(t) \, \longrightarrow \, S(t) + \varepsilon \kappa(t)
    \end{split}
\end{equation}
from which it follows that 
\begin{equation*}
    \delta\mathcal{Z} = \int_{t_1}^{t_2} dt\;\left[\,\varepsilon(1+\lambda)\left(\frac{\partial\mathcal{L}^H}{\partial q}\eta + \frac{\partial\mathcal{L}^H}{\partial \dot{q}}\dot{\eta} + \frac{\partial\mathcal{L}^H}{\partial \ddot{q}}\ddot{\eta} + \,\cdots\, + \frac{\partial \mathcal{L}^H}{\partial S}\kappa \right) - \varepsilon\lambda\dot{\kappa}\right]
\end{equation*}
where the dots $\cdots$ represent the remaining terms of the type $\frac{\partial\mathcal{L}^H}{\partial q_{\alpha}}\eta^{(\alpha)}$, up to order $k$. Integrating by parts, and discarding boundary terms, we find that
\begin{equation}\label{Eq:ExtremisatipnConditions}
    \begin{split}
        (1+\lambda)\frac{\partial\mathcal{L}^H}{\partial q} - \frac{d}{dt}\left[(1+\lambda)\frac{\partial\mathcal{L}^H}{\partial \dot{q}}\right] + \frac{d^2}{dt^2}\left[(1+\lambda)\frac{\partial\mathcal{L}^H}{\partial \ddot{q}}\right] -\,\cdots\, = \,0\\
        \frac{d\lambda}{dt} = -\,(1+\lambda)\frac{\partial\mathcal{L}^H}{\partial S}
    \end{split}
\end{equation}
The second of these equations may then be used to calculate higher derivatives of $\lambda$, with which we may eliminate all reference to this Lagrange multiplier in the first. It is illustrative to group terms in the resulting expression as follows
\begin{equation*}
    \frac{\partial\mathcal{L}^H}{\partial q} - \left(\frac{d}{dt}\frac{\partial\mathcal{L}^H}{\partial \dot{q}} + \frac{\partial\mathcal{L}^H}{\partial \dot{q}}\frac{\partial\mathcal{L}^H}{\partial S}\right) + \left(\frac{d^2}{dt^2}\frac{\partial\mathcal{L}^H}{\partial \ddot{q}} - \frac{\partial\mathcal{L}^H}{\partial \dot{q}}\frac{d}{dt}\frac{\partial\mathcal{L}^H}{\partial S} -2\frac{\partial\mathcal{L}^H}{\partial S}\frac{d}{dt}\frac{\partial\mathcal{L}^H}{\partial \ddot{q}} + \frac{\partial\mathcal{L}^H}{\partial \ddot{q}}\left(\frac{\partial\mathcal{L}^H}{\partial S}\right)^2\right) - \,\cdots\,=0
\end{equation*}
The reason we have chosen to collect the terms in this manner, is that we can see that the first bracket is precisely the Lagrangian total derivative $D_{\mathcal{L}}$ acting on $\frac{\partial\mathcal{L}^H}{\partial \dot{q}}$, where the $-\,\mathcal{L}\frac{\partial}{\partial S}$ from (\ref{Eq:Lagtot}) is absent, since $\frac{\partial\mathcal{L}^H}{\partial \dot{q}}$ is independent of the action. Additionally, the second bracket coincides with the action of the composition $D_{\mathcal{L}}\circ D_{\mathcal{L}}$ on $\frac{\partial\mathcal{L}^H}{\partial \ddot{q}}$. A similar analysis for higher - order terms confirms the general expression
\begin{equation}
    \sum_{\alpha=0}^k(-1)^{\alpha}D^{\alpha}_{\mathcal{L}}\left(\frac{\partial\mathcal{L}^H}{\partial q_{\alpha}}\right) = 0
\end{equation}
Finally, it should be evident that, in the case of multiple generalised coordinates $q^i$, we should consider separate variations $q^i(t)\rightarrow q^i(t)+\varepsilon \eta^i(t)$. Extremisation of the action in this case leads to an expression which coincides perfectly with (\ref{Eq:contactEL2}).
\section{Higher - Order FLRW System with General Lapse}\label{Appendix:B}
Our goal in this appendix is to outline the details of the calculation carried out in section (\ref{Subsec:EgFLRW}), when a general lapse function $N(t)$ is employed. We shall continue to impose Dirichlet conditions on the spacetime boundary $\partial M$, such that the additional modified GHY surface term vanishes. Retaining a non - trivial lapse function in (\ref{Eq:Ricci}), the resulting Lagrangian density, which generalises (\ref{Eq:FLRWlagrangian}), is given by
\begin{equation}\label{Eq:BFLRWLagrangian}
    \mathcal{L} = -\,\frac{1}{24\pi G}\frac{\dot{v}^2}{vN}-\frac{\lambda}{4\pi G}\left(\frac{\ddot{v}^2}{vN^3}-\frac{\dot{v}^4}{3v^3N^3} - \frac{2\dot{v}\ddot{v}\dot{N}}{vN^4} + \frac{\dot{v}^2\dot{N}^2}{vN^5}\right) + vN\left(\frac{1}{2}\ddot{\varphi}^2 +\frac{1}{2}\dot{\varphi}^2 - V(\varphi)\right)
\end{equation}
The introduction of a new lapse does not affect the properties of $\mathcal{L}$ under rescaling, and we find that the vector field (\ref{Eq:FLRWscaling}) is still a scaling symmetry of degree one. Consequently, we may proceed to introduce the change of coordinate $v = e^{\rho}$, after which the Lagrangian now reads
\begin{equation}\label{Eq:BFLRWRhoLagrangian}
    \begin{split}
        \mathcal{L} = -\,e^{\rho} \left[ \frac{1}{24\pi G} \frac{\dot{\rho}^2}{N} + \frac{\lambda}{4\pi G}\left(\frac{\dot{N}^2\dot{\rho}^2}{N^5} - \frac{2\dot{N}\dot{\rho}(\dot{\rho}^2+\ddot{\rho})}{N^4} + \frac{2\dot{\rho}^2(\dot{\rho}^2+ 3\ddot{\rho}) + 3\ddot{\rho}^2}{3N^3}\right) \right. \\
        \left. - N \left(\frac{1}{2}\ddot{\varphi}^2 +\frac{1}{2}\dot{\varphi}^2 - V(\varphi)\right)\right]
    \end{split}
\end{equation}
Following identical steps to those described in the main text, we deduce the following Herglotz Lagrangian
\begin{equation}\label{Eq:BFLRWHerglotzLagrangian}
    \begin{split}
        \mathcal{L}^H = -\, \frac{1}{24\pi G} \frac{\chi^2}{N} - \frac{\lambda}{4\pi G}\left(\frac{\dot{N}^2\chi^2}{N^5} - \frac{2\dot{N}\chi(\chi^2+\dot{\chi})}{N^4} + \frac{2\chi^2(\chi^2+ 3\dot{\chi}) + 3\dot{\chi}^2}{3N^3}\right) \\
        + N \left(\frac{1}{2}\ddot{\varphi}^2 +\frac{1}{2}\dot{\varphi}^2 - V(\varphi)\right) - \chi S
    \end{split}
\end{equation}
Recall from our general formalism that the equations of motion are deduced imposing that $\dot{S}=\mathcal{L}^H$; in this case, the relevant action, calculated from (\ref{Eq:BFLRWRhoLagrangian}), is given by
\begin{equation}\label{Eq:BFLRWAction}
    S = -\frac{1}{12\pi G}\frac{\dot{\rho}}{N} + \frac{\lambda}{2\pi G}\left(\frac{3\dot{N}^2\dot{\rho}}{N^5} - \frac{\ddot{N}\dot{\rho}+ \dot{N}(3\ddot{\rho}+\dot{\rho}^2)}{N^4} + \frac{3\dddot{\rho}-\dot{\rho}^3 + 3\ddot{\rho}\dot{\rho}}{3N^3} \right)
\end{equation}
Consequently, we find that the dynamical evolution of the variable $\chi$ is governed by
\begin{equation}\label{Eq:BFLRWChiEoM}
    \begin{split}
        \dddot{\chi} = \frac{\dddot{N}\chi + 2\ddot{N}(2\dot{\chi} + \chi^2) + \dot{N}(6\ddot{\chi}+ 9\chi\dot{\chi} + \chi^3)}{N} - \frac{7\dot{N}^2(2\dot{\chi}+ \chi^2) - 20\chi\dot{N}\ddot{N}}{2N^2} + \frac{15\chi \dot{N}^3}{N^3} - \frac{\chi N \dot{N}}{3\lambda}\\
        + \frac{2\pi G N^4}{\lambda}\left(\frac{1}{2}\ddot{\varphi}^2 + \frac{1}{2}\dot{\varphi}^2 - V(\varphi)\right) - 2\chi\ddot{\chi} - \frac{3}{2}\dot{\chi}^2 - \chi^2\dot{\chi} + \frac{N^2}{12\lambda}\left( 2\dot{\chi} + \chi^2 \right)
    \end{split}
\end{equation}
which is the generalisation of (\ref{Eq:FLRWHerglotzEoM}), for non - trivial lapse. The Jacobi - Ostrogradsky momenta associated with the Herglotz Lagrangian receive substantial modification due to the presence of the lapse function; in particular, we find that
\begin{equation*}
    \widehat{\Pi}^{0}_{\chi} = \frac{\partial \mathcal{L}^H}{\partial \dot{\chi}} = \frac{\lambda}{2\pi G N^3}\left[ \frac{\dot{N}\chi}{N} - (\chi^2+2\dot{\chi}) \right]
\end{equation*}
\vspace{-5mm}
\begin{equation*}
        \widehat{\Pi}^1_{\varphi} = \frac{\partial \mathcal{L}^H}{\partial \ddot{\varphi}} = N\ddot{\varphi}\quad\quad\quad\quad
        \widehat{\Pi}^0_{\varphi} = \frac{\partial \mathcal{L}^H}{\partial \dot{\varphi}} - d_T (\widehat{\Pi}^1_{\varphi})=N\dot{\varphi} - N\dddot{\varphi} - \dot{N}\ddot{\varphi}
\end{equation*}
From here, it is then straightforward to continue the remainder of the calculation, obtaining the contact Hamiltonian and its associated equations of motion; we leave this as an exercise to the reader.
\section*{Acknowledgement}
We would like to thank the anonymous referee, whose valuable feedback contributed to the revision and improvement of this manuscript.
\bibliographystyle{unsrt}
\bibliography{refs}
\end{document}